\documentclass[aps,twocolumn,superscriptaddress,showpacs]{revtex4-1}
\usepackage[T1]{fontenc}
\usepackage{graphicx}
\usepackage{bm}
\usepackage{dcolumn}

\usepackage{epstopdf}
\usepackage{ulem}
\usepackage{siunitx}
\usepackage[ngerman]{babel}
\addto\captionsngerman{}  
\addto\captionsngerman{}  
\usepackage{natbib}
\usepackage{tabularx}
\newcolumntype{L}[1]{>{\raggedright\arraybackslash}p{#1}}
\newcolumntype{C}[1]{>{\centering\arraybackslash}p{#1}}
\newcolumntype{R}[1]{>{\raggedleft\arraybackslash}p{#1}}

\usepackage{array}
\usepackage{textgreek}
\usepackage{upgreek}
\usepackage[dvipsnames]{xcolor}

\usepackage{multirow}

\begin{document}

\bibliographystyle{apsrev}

\title{Lattice dynamics of endotaxial silicide nanowires}

\author{J. Kalt}
 \affiliation{Laboratory for Applications of Synchrotron Radiation, Karlsruhe Institute of Technology, \textit{D-76131} Karlsruhe, Germany}
 \affiliation{Institute for Photon Science and Synchrotron Radiation, Karlsruhe Institute of Technology, \textit{D-76344} Eggenstein-Leopoldshafen, Germany}

\author{M. Sternik}
 \affiliation{Institute of Nuclear Physics, Polish Academy of Sciences, \textit{PL-31342} Krak\'{o}w, Poland}

\author{B. Krause}
 \affiliation{Institute for Photon Science and Synchrotron Radiation, Karlsruhe Institute of Technology, \textit{D-76344} Eggenstein-Leopoldshafen, Germany}

\author{I. Sergueev}
 \affiliation{Deutsches Elektronen-Synchrotron, \textit{D-22607} Hamburg, Germany}

\author{M. Mikolasek}
 \affiliation{ESRF - The European Synchrotron, \textit{F-38000} Grenoble, France}

\author{D. Merkel}
 \affiliation{Institute for Particle and Nuclear Physics, Wigner Research Centre for Physics, Hungarian Academy of Sciences, \textit{H-1525} Budapest, Hungary}
 \affiliation{Centre for Energy Research, \textit{H-1121} Budapest, Hungary}

\author{D. Bessas}
 \affiliation{ESRF - The European Synchrotron, \textit{F-38000} Grenoble, France}

\author{O. Sikora}
 \affiliation{Institute of Nuclear Physics, Polish Academy of Sciences, \textit{PL-31342} Krak\'{o}w, Poland}

\author{T. Vitova}
 \affiliation{Institute for Nuclear Waste Disposal, Karlsruhe Institute of Technology, \textit{D-76344} Eggenstein-Leopoldshafen, Germany}

\author{J. Göttlicher}
 \affiliation{Institute for Photon Science and Synchrotron Radiation, Karlsruhe Institute of Technology, \textit{D-76344} Eggenstein-Leopoldshafen, Germany}

\author{R. Steininger}
 \affiliation{Institute for Photon Science and Synchrotron Radiation, Karlsruhe Institute of Technology, \textit{D-76344} Eggenstein-Leopoldshafen, Germany}

\author{P. T. Jochym}
 \affiliation{Institute of Nuclear Physics, Polish Academy of Sciences, \textit{PL-31342} Krak\'{o}w, Poland}

\author{A. Ptok}
 \affiliation{Institute of Nuclear Physics, Polish Academy of Sciences, \textit{PL-31342} Krak\'{o}w, Poland}

\author{O. Leupold}
 \affiliation{Deutsches Elektronen-Synchrotron, \textit{D-22607} Hamburg, Germany}

\author{H.-C. Wille}
 \affiliation{Deutsches Elektronen-Synchrotron, \textit{D-22607} Hamburg, Germany}
 
 \author{A. I. Chumakov}
 \affiliation{ESRF - The European Synchrotron, \textit{F-38000} Grenoble, France}
 
\author{P. Piekarz}
 \affiliation{Institute of Nuclear Physics, Polish Academy of Sciences, \textit{PL-31342} Krak\'{o}w, Poland}

\author{K. Parlinski}
 \affiliation{Institute of Nuclear Physics, Polish Academy of Sciences, \textit{PL-31342} Krak\'{o}w, Poland}
 
 \author{T. Baumbach}
 \affiliation{Laboratory for Applications of Synchrotron Radiation, Karlsruhe Institute of Technology, \textit{D-76131} Karlsruhe, Germany}
 \affiliation{Institute for Photon Science and Synchrotron Radiation, Karlsruhe Institute of Technology, \textit{D-76344} Eggenstein-Leopoldshafen, Germany}

\author{S. Stankov}
 \email{svetoslav.stankov@kit.edu}
 \affiliation{Laboratory for Applications of Synchrotron Radiation, Karlsruhe Institute of Technology, \textit{D-76131} Karlsruhe, Germany}
 \affiliation{Institute for Photon Science and Synchrotron Radiation, Karlsruhe Institute of Technology, \textit{D-76344} Eggenstein-Leopoldshafen, Germany}

\date{\today}

\begin{abstract}
Self-organized silicide nanowires are considered as main building blocks of future nanoelectronics and have been intensively investigated. 
In nanostructures, the lattice vibrational waves (phonons) deviate drastically from those in bulk crystals, which gives rise to anomalies in thermodynamic, elastic, electronic, and magnetic properties.
Hence, a thorough understanding of the physical properties of these materials requires a comprehensive investigation of the lattice dynamics as a function of the nanowire size.
We performed a systematic lattice dynamics study of endotaxial FeSi$_2$ nanowires, forming the metastable, surface-stabilized $\upalpha$-phase, which are \textit{in-plane} embedded \textit{into} the Si(110) surface. 
The average widths of the nanowires ranged from 24 to 3\,nm, their lengths ranged from several $\upmu$m to about 100\,nm.
The Fe-partial phonon density of states, obtained by nuclear inelastic scattering, exhibits a broadening of the spectral features with decreasing nanowire width.
The experimental data obtained along and across the nanowires unveiled a pronounced vibrational anisotropy that originates from the specific orientation of the tetragonal $\upalpha$-FeSi$_2$ unit cell on the Si(110) surface.
The results from first-principles calculations are fully consistent with the experimental data and allow for a comprehensive understanding of the lattice dynamics of endotaxial silicide nanowires.
\end{abstract}

\maketitle

\section{Introduction}\label{Introduction}

Metallic silicides constitute an important part of current microelectronics, serving as Schottky barriers and ohmic contacts, gate electrodes, local interconnects and diffusion barriers \cite{Murarka_silicides_microelectronics,Chen_book,chen_silicides_microelectronics}. 
The enormous degree of miniaturization of nowadays integrated circuits imposes severe restrictions on the spatial dimensions of these components.
New materials and configurations are constantly researched for nanoelectronic applications and endotaxial silicide nanowires (NWs), self-organized on the Si surface, have been considered as promising candidates \cite{endotaxial_review}.
The endotaxial mechanism implies the formation of  \textit{in-plane}  unidirectionally aligned, high-aspect ratio NWs grown partially \textit{into} the substrate \cite{endotaxial_nw_prl}.
These nanostructures are readily integrated with Si technology and exhibit high crystal-phase purity and thermal stability, sharp interfaces and  Schottky barrier heights which are tunable by the choice of silicide material \cite{nanotechnology}.
FeSi$_2$ is a particularly attractive silicide since it exhibits several crystal phases, namely the room-temperature stable semiconducting $\upbeta$-phase, high-temperature metallic $\upalpha$-phase and surface-stabilized metallic $s$- and $\upgamma$-phases \cite{endotaxial_review}.
Due to the very small lattice mismatch with certain crystallographic planes of Si, the tetragonal $\upalpha$-FeSi$_2$  can also be stabilized at room temperature in epitaxial nanostructures on Si surfaces.
However, the reports about the crystal structure of FeSi$_2$ NWs formed on Si(110) remain contradictory, spanning cubic ($s$ or $\upgamma$) \cite{cubic1,cubic2,cubic3} and tetragonal ($\upalpha$) \cite{tetra} phases.

When the dimensions of nanostructures approach the characteristic phonon mean free paths (from several nanometers up to micrometers), the phonon dispersions and the phonon density of states (PDOS), which characterize the lattice dynamics of a material, begin to deviate from those of the bulk counterparts.
These deviations imply significant modifications of thermodynamic and elastic properties, which are directly related to the lattice dynamics, such as lattice heat capacity, vibrational entropy, mean force constants, sound velocity and thermal conductivity. 
They also lead to an enhanced electron-phonon, spin-phonon and phonon-phonon scattering at surfaces and interfaces \cite{Bozyigit}.
In conjunction with a possible emergence of phonon quantum phenomena at very small dimensions \cite{hepplestone_apl_2005}, these effects could significantly deteriorate the electron and spin transport in nanoscale interconnects \cite{nanowires_resis,nanowires_phonon,tobler_resistivity}. 

The phonon dispersions and PDOS of 1D nanostructures have been subjects of intensive theoretical studies, predicting features that differ significantly from those in the 3D counterparts. 
The most prominent effects are confined bands and band gaps, acoustic modes with non-zero frequencies at $q=0$ ($q$ is the phonon wave vector), non-linear dispersion for small $q$ and a complex displacement field \cite{theor1,theor2,theor3,theor4,theor5,theor7,theor8,theor10,theor11}.
Consequently, anomalies in thermal conductivity \cite{therm0,therm1,therm3,therm4,therm7,therm8,therm10,therm12} and electron-phonon interactions \cite{eph1,eph5,eph6,eph7} were predicted and strategies for their tailoring in NWs were suggested \cite{maldovan3}.

Unlike the intensive theoretical studies, the experimental reports on lattice dynamics of NWs remain scarce.
Optical phonon confinement phenomena in Si \cite{zhang,wang,piscanec,adu,dhara} and III-V \cite{luca} NWs  were studied by Raman spectroscopy.
Resonant and propagative coherent acoustic phonon modes were investigated by time-resolved spectroscopy with visible light \cite{perrin} and x-rays \cite{mariager}.
Using Brillouin-Mandelstam light scattering spectroscopy, surprisingly strong confinement effects in the acoustic phonon dispersions were observed in individual GaAs NWs with diameters exceeding by an order of magnitude the phonon mean-free path \cite{balandin3}.
Employing nuclear inelastic scattering (NIS) on the $^{125}$Te isotope, the Te-partial PDOS of Bi$_2$Te$_3$ NWs array with an average NW diameter of 56\,nm was determined, unveiling a reduction of the speed of sound by 7\,\% compared to the bulk material \cite{bessas}. 
By application of the same experimental technique on the $^{119}$Sn resonance, a correlation between the lattice softening in Sn NWs with a diameter between 100 and 18\,nm and an increase of the critical temperature of the superconducting state was established \cite{temst}.
Despite their  potential applications in nanoelectronics, the lattice dynamics of endotaxial silicide NWs remains unexplored.

Here we present a systematic lattice dynamics study of endotaxial FeSi$_2$ NWs formed on Si(110) for a large range of sizes. 
The Fe-partial PDOS exhibits a broadening of the spectral features with decreasing NW width. 
The experimental data obtained along and across the NWs unveil a pronounced vibrational anisotropy that originates from the specific orientation of the tetragonal $\upalpha$-FeSi$_2$ unit cell on the Si(110) surface and is fully consistent with the results from first-principles calculations.

\section{Experimental and theoretical details}\label{EXPERIMENT}

Endotaxial FeSi$_2$ NWs were grown on the 16\,$\times$\,2 reconstructed  Si(110) surface under ultra high vacuum (UHV) conditions (P$<$1$\times$10$^{-8}$\,Pa). 
The substrates were degassed in UHV at 650$\,^\circ$C for 4\,h, followed by the removal of the native SiO$_2$ layer by heating two times to 1250$\,^\circ$C for 30 seconds. 
Subsequently, the Si(110) surface was stabilized at the growth temperature $T_G$ and  a certain amount $\theta_{Fe}$ of high purity iron, enriched to 96\,\% in the M\"ossbauer-active isotope $^{57}$Fe, was deposited.
The coverage was controlled by a quartz oscillator with an accuracy of 10\,\% and is given  in monolayer (ML) units, with 1\,ML\,=\,4.78\,$\times$\,10$^{14}$ Fe atoms/cm$^2$. 
Details of the growth and experimental conditions used for the investigated samples, hereinafter referred to as S1-S7, are summarized in Table\,\ref{tab:samples}.  
All measurements described in the following were conducted at room temperature. 
The crystal structure of the samples was investigated with reflection high-energy electron diffraction (RHEED), the surface topography was determined by non-contact atomic force microscopy (AFM) in an {\sc Omicron Large Sample} scanning probe microscope connected to the UHV-cluster.
Samples S1, S2, S3, S5 and S6 were subsequently capped with 4\,nm of amorphous Si sputtered at room temperature in a chamber \cite{sputter_krause} with a base pressure of P$<$1$\times$10$^{-6}$\,Pa also  connected to the UHV-cluster. 
The flux of the sputter gas Ar was 0.8\,sccm, corresponding to a pressure of 0.36 Pa.

The local crystal structure of the NWs in S2, S6 and S7 was investigated by Fe $K$-edge x-ray absorption spectroscopy at the SUL-X beamline of the synchrotron radiation source KARA at KIT.
After calibration with an $\upalpha$-Fe metal foil to the Fe\,$K$-edge at 7112\,eV,  the fluorescence emission of the samples was recorded up to $k\,=\,14\,$\AA$^{-1}$. 
The incoming x-ray beam was parallel to Si[$\bar{1}10$], i.e. oriented along the NWs. 
A beam-to-sample-to-detector geometry of 45$^{\circ}$/45$^{\circ}$ was applied, using a collimated x-ray beam of about 0.8\,mm\,$\times$\,0.8\,mm, or focused x-ray beam with  0.35\,mm\,$\times$\,0.15\,mm (h\,$\times$\,v, FWHM) at the sample position. 
The obtained extended x-ray absorption fine structure (EXAFS) spectra were processed with the {\sc ATHENA} and {\sc ARTEMIS} programs included in the {\sc IFEFFIT} package \cite{ravel_athena}.
The spectra were weighted by $k^1$, $k^2$, $k^3$ within a $k$-range of  3.8\,-\,13.2\,\AA$^{-1}$.
The data was modeled in the real space with Hanning windows and $dk$\,=\,2 within a range of 1.0\,-\,2.7\,\AA~ using a shell-by-shell approach.
Multiple scattering paths do not contribute in the modeled R region.
The single scattering paths were calculated with {\sc FEFF6} for the crystal structure of $\alpha$-FeSi$_2$.
The amplitude reduction factor was set to 0.7 and was fixed during the fitting process. 
It was obtained by modeling the EXAFS spectrum of the $\alpha$-Fe foil used for calibration. 
The Debye-Waller factors of Si were variable fit parameters, whereas for Fe they were fixed to the values obtained from the NIS experiment.

\begin{table}[b]
 \caption{Overview of the investigated samples. $\theta_{Fe}$ stands for the deposited amount of $^{57}$Fe, $T_G$  for the growth temperature and $\bar{w}$ for the average NW width. The last column denotes if the sample was capped with Si or measured \textit{in situ} during the NIS experiment.  
} 
\renewcommand{\arraystretch}{1.1}
   \begin{tabular*}{0.48\textwidth}{@{\extracolsep{\fill}}ccccc}
\hline
\hline
Sample   	 & $\theta_{Fe}$ [ML]  			&    $T_G$  [$\,^\circ$C]    		&	$\bar{w}$ [nm]	& NIS exp. 			\\                  
\hline
S1	    	 &   3\,$\pm$\,0.3     			&     825\,$\pm$\,20		   		&	24\,$\pm$\,7	& Si cap 			\\ 
S2	    	 &   6\,$\pm$\,0.6				&     700\,$\pm$\,10		   		&	18\,$\pm$\,5	& Si cap 			\\ 
S3	 	     &   2\,$\pm$\,0.2				&     700\,$\pm$\,10				&	10\,$\pm$\,3	& Si cap   			\\ 
S4	         &   2\,$\pm$\,0.2				&     700\,$\pm$\,10			   	&	11\,$\pm$\,3	& \textit{in situ}	\\ 
S5		 	 &   1.5\,$\pm$\,0.2~~~	    	&     600\,$\pm$\,10   				&	~~4\,$\pm$\,1	& Si cap 		  	\\ 
S6		 	 &   1\,$\pm$\,0.1	    		&     600\,$\pm$\,10   				&	~~3\,$\pm$\,1	& Si cap 		  	\\ 
S7	     	 &   4\,$\pm$\,0.3     			&     825\,$\pm$\,20		   		&	26\,$\pm$\,7	& - 				\\ 
\hline
\hline	
 \end{tabular*}
 \label{tab:samples}
\end{table}

The Fe-partial PDOS was obtained \cite{evaluation_pdos} by NIS experiments \cite{Seto_PRL_NIS,Sturhahn_PRL_NIS} performed at the Dynamics Beamline P01 \cite{p01} at PETRA III and the Nuclear Resonance Beamline ID18 \cite{id18} at the ESRF. 
At both beamlines the measurements were performed in grazing-incidence geometry with an incidence angle\,<\,0.2$^{\circ}$ and an x-ray beam with dimensions of 1.5\,mm\,$\times$\,0.01 mm (h\,$\times$\,v, FWHM).
The energy dependence of the probability for nuclear inelastic absorption was measured by tuning the energy of the x-ray beam around the $^{57}$Fe resonance at 14.413\,keV  with an energy resolution of 0.7\,meV for S1, S2 (ID18), 1.0\,meV for S3, S4 (P01) and 1.3\,meV for S5, S6 (P01).
Sample S4 was transported to the beamline  and measured under UHV condition (P$<$5$\times$10$^{-7}$\,Pa) in a dedicated chamber \cite{ibrahimkutty_chamber}.

\begin{figure}[t!]
 \centering
 \includegraphics[width=1\columnwidth]{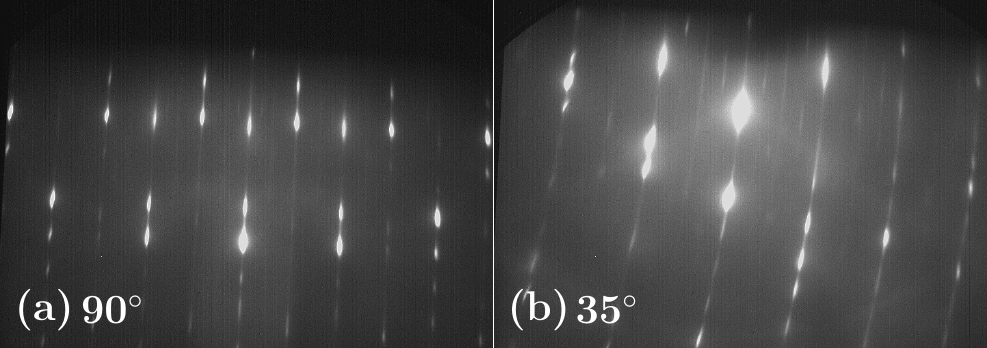}
 \caption{RHEED patterns of sample S2 obtained with E\,=\,28\,keV (a) perpendicular to the NWs (along $\upalpha$-FeSi$_2[ 44\bar{1}]$) and (b) at an angle of 35$^\circ$ between electron beam and NWs.
 }
 \label{fig:RHEED}
\end{figure}

First-principles calculations were performed within the density functional theory implemented in the VASP code \cite{vasp1,vasp2},  employing the generalized gradient approximation \cite{PBE1,PBE2}.
The phonon dispersions and PDOS were calculated using the direct method \cite{phonon1} incorporated into the PHONON program  \cite{phonon2}.
To account for the tensile epitaxial strain between the FeSi$_2$ crystal and the Si(110) surface, the calculations were done for 0.5\,\% tensile-strained  $\upalpha$-phase FeSi$_2$ ($a\,=\,2.716$\,\AA, 
$c\,=\,5.166$\,\AA). 
Further details are given elsewhere \cite{kalt_alpha_islands}. 

To ensure a valid comparison of the parameters obtained by fitting the experimental data with the \textit{ab initio} calculated PDOS, the energy resolution each sample was measured with was considered. 
The \textit{ab initio} calculated energy dependence of the probability for nuclear inelastic absorption was convoluted with a Voigt profile with the FWHM corresponding to the energy resolution used for the respective sample. 
Subsequently, the PDOS used for the fitting process of each sample was calculated.

\section{Results and discussion}\label{Results}

\subsection{Structural investigation}

Figure\,\,\ref{fig:RHEED} shows generic RHEED patterns of NWs obtained with the wave vector of the electron beam being oriented (a) 90$^\circ$ and (b) 35$^\circ$ with respect to the NWs.
At 90$^\circ$  the pattern is composed of several diffraction spots superimposed on straight streaks. 
When the angle between the NWs and the wave vector of the electron beam is reduced to 35$^\circ$, the streaks are bended and the diffraction spots follow their curvature. 
This observation is explained by the reciprocal space planes of one-dimensional atomic chains with high crystalline order along the chain orientation \cite{1d_RHEED_theo,dobson1982_GaAs_domain,1d_RHEED_exp} and confirms the formation of single-crystalline, unidirectionally aligned NWs.
The RHEED images obtained for S1-S6 revealed the same pattern throughout the entire range of growth parameters, indicating that for all samples the NWs exhibit the same crystal structure.
A detailed discussion of the RHEED results is given in the Supplemental Material \cite{sm}.

\begin{figure}[t]
 \centering
      {\includegraphics[width= 0.49\textwidth]{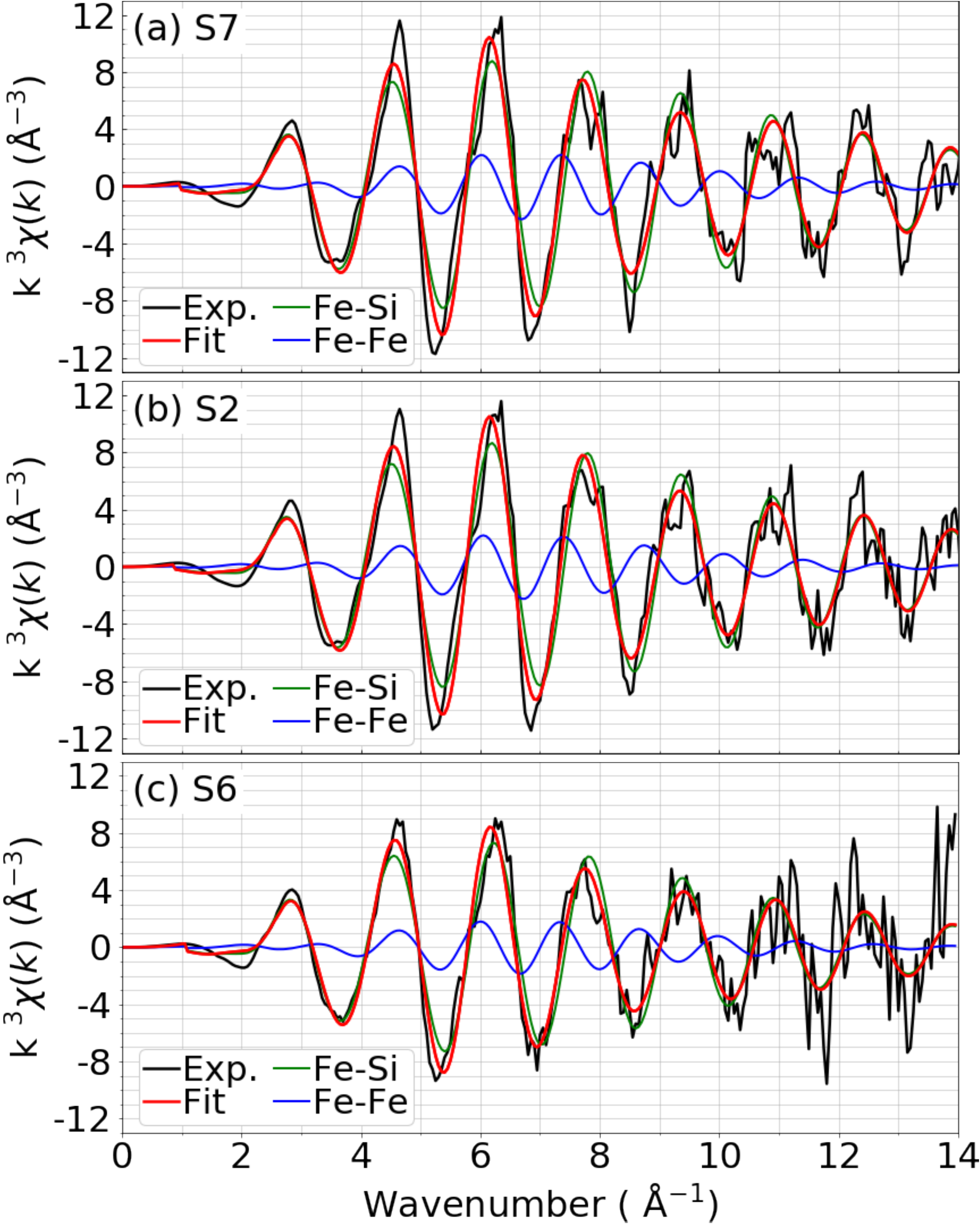}}\hfill  
\caption{ Fe\,$K$-edge EXAFS spectra of (a) S7, (b) S2 and (c)  S6,  compared with the respective best fit results and Fe-Fe, Fe-Si scattering path subspectra.
For the modeling the $\alpha$-FeSi$_2$ crystal structure was assumed.
}
 \label{fig:EXAFS}
  \end{figure}

\begin{table}[t]
\begin{center}
\caption{
Debye-Waller factor ($\sigma^2$), coordination numbers, and interatomic distances (\textit{d}) obtained from modeling of the experimental EXAFS spectra  and theoretical values for the expected FeSi$_2$ phases. 
The coordination numbers were calculated with the x-ray beam  projected along the respective crystal direction parallel to Si[$\bar{1}10$], i.e. along  $\upalpha$-FeSi$_2$[$1\bar{1}0$], $\upbeta$-FeSi$_2$[$010$], $\upgamma$-FeSi$_2$[$\bar{1}10$] and s-FeSi$_2$[$\bar{1}10$].
The $\sigma^2$ values for the Fe-Si scattering path are derived from the modeling of the experimental EXAFS spectra, for the Fe-Fe scattering path they were fixed to the mean square displacement values obtained from the NIS experiments.
The theoretical values for $\alpha$- and $\beta$-phase are obtained from ICSD 5257 and 9119, respectively, for  s- and $\gamma$-phase no database values available.
} 
\renewcommand{\arraystretch}{1.05}
   \begin{tabular*}{0.48\textwidth}{@{\extracolsep{\fill}}ccccc}
\hline
\hline\\[-3.5mm]
&	\shortstack{scattering\\path}	&   $\sigma^2$  (10$^{-2}$\AA $^2$) & \shortstack{coord.\\number}     & \vspace{0.01mm}  \textit{d} (\AA)		\\
\hline
\multirow{2}{*}{S7}	 					
& Fe-Si 	& 0.41\,$\pm$\,0.05	&			7.4\,$\pm$\,0.4					&				2.36\,$\pm$\,0.01				\\
& Fe-Fe		& 1.00\,$\pm$\,0.02	&			2.6\,$\pm$\,0.3					&				2.68\,$\pm$\,0.01			\\
\hline
\multirow{2}{*} {S2}	 				
& Fe-Si     &	0.41\,$\pm$\,0.04 	&			7.3\,$\pm$\,0.3				&				2.35\,$\pm$\,0.01		\\
& Fe-Fe		&	1.01\,$\pm$\,0.02	&			2.8\,$\pm$\,0.3				&				2.67\,$\pm$\,0.01		\\
\hline
\multirow{2}{*} {S6}	 			
& Fe-Si 	&	0.52\,$\pm$\,0.08	&			6.6\,$\pm$\,0.5				&				2.35\,$\pm$\,0.01		\\
& Fe-Fe		&	1.05\,$\pm$\,0.02	&			2.3\,$\pm$\,0.5				&				2.69\,$\pm$\,0.01			\\
\hline
\multirow{2}{*} {$\alpha$-phase}					
& Fe-Si 	& - &			 8    					&			2.36	\\
& Fe-Fe		& - &			 2    					&			2.70  \\
\hline
\multirow{2}{*} {$\beta$-phase}						
& Fe-Si 	& - &			 6    					&			2.36\\
& Fe-Fe		& - &			 2    					&			2.97\\
\hline
\multirow{2}{*} {s-phase}					
& Fe-Si 	& - &			 4    					&			2.39\\
& Fe-Fe		& - &			 4    					&			2.76\\
\hline
\multirow{2}{*} {$\gamma$-phase}						
& Fe-Si 	& - &			 4    					&			2.33\\
& Fe-Fe		& - &			 10   					&			3.81\\
\hline
\hline
\end{tabular*}
 \label{tab:tab_EXAFS}
\end{center}
\end{table}

In Fig. \ref{fig:EXAFS} the  experimental EXAFS spectra in $k$ space of samples S7, S2 and S6 are compared to the respective best fit results. 
Since S7 was grown at very similar conditions as S1 (see Table\,1), it is concluded that the NWs of these two samples exhibit the same crystal structure. 
In Table \ref{tab:tab_EXAFS} the interatomic distances and coordination numbers of the Si and Fe nearest neighbors, obtained by modeling of the experimental data,  are compared with the values theoretically predicted for the FeSi$_2$ phases formed on Si surfaces. 
For the determination of the theoretical values it has to be considered that in single crystals the intensity of the EXAFS signal depends on the orientation of the incoming x-ray beam relative to the crystal axes.
All spectra were  measured with the wave vector of the x-ray beam being parallel to the NWs, i.e. oriented along Si[$\bar{1}10$].
The crystal directions being parallel to Si[$\bar{1}10$] are:  $\upalpha$-FeSi$_2$[$1\bar{1}0$], $\upbeta$-FeSi$_2$[$010$], $\upgamma$-FeSi$_2$[$\bar{1}10$] and s-FeSi$_2$[$\bar{1}10$].
Correspondingly, the coordination numbers given in Table \ref{tab:tab_EXAFS} were calculated with the x-ray beam  projected along the respective FeSi$_2$ crystal direction.
The results for both parameters exclude the formation of $\upbeta$-, $s$- or $\upgamma$-FeSi$_2$ and, in agreement with an earlier report \cite{tetra}, reveal that the investigated NWs  exhibit the tetragonal $\alpha$-phase.
Furthermore, the values obtained from the modeling of the EXAFS data show an increase of the Debye-Waller factor of the Si atoms in the smallest wires (S6).
The fit results for the interatomic distances  do not show a size-dependent behavior, while the coordination numbers for Fe-Si and Fe-Fe are reduced in S6 compared to S7 and S2. 
The reason for this is the increased interface-to-volume ratio in the smallest NWs of S6 compared to S7 and S2.

\begin{figure}[t]
\includegraphics[width=0.99\columnwidth]{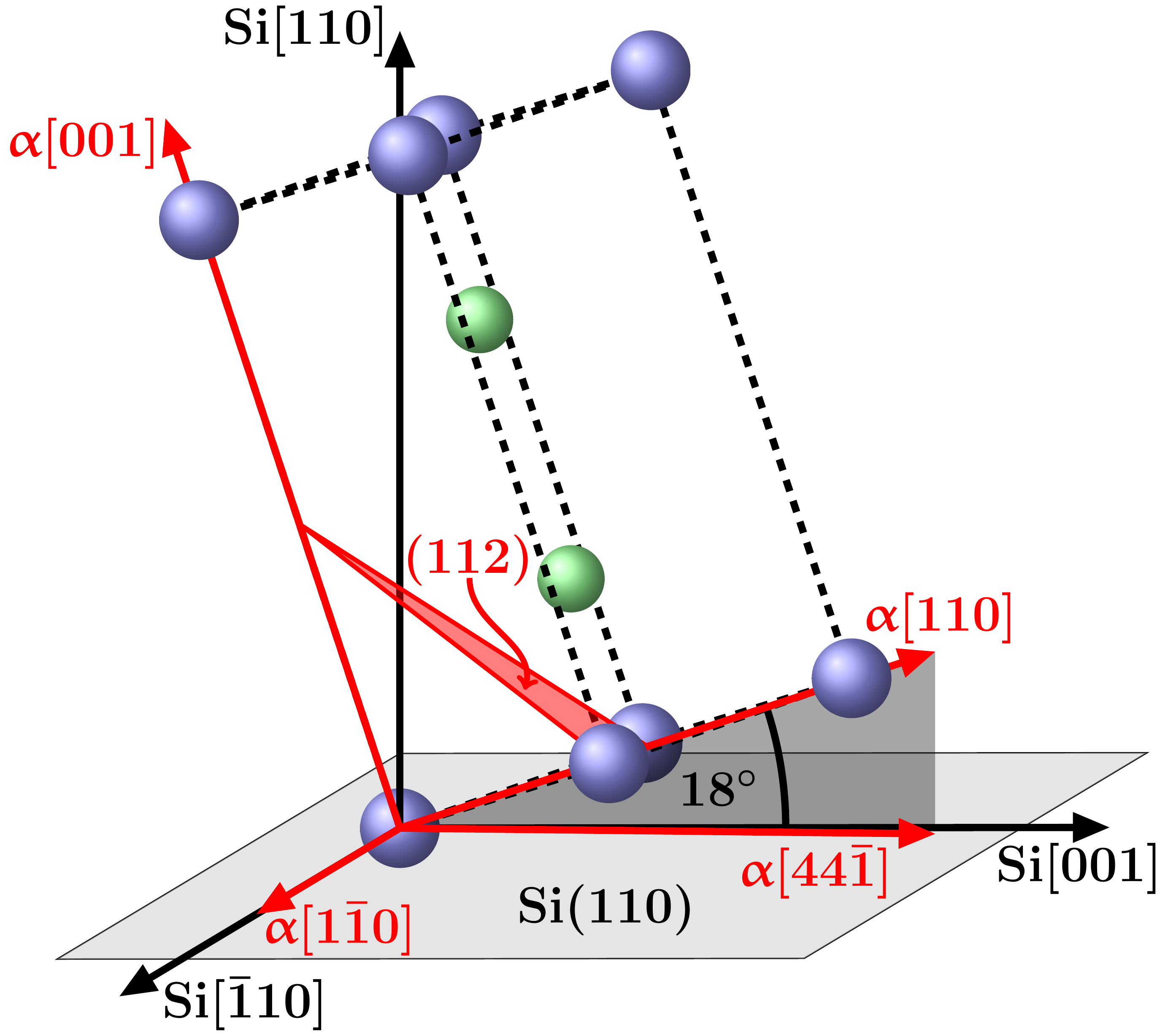}
\caption{Orientation  of the $\upalpha$-FeSi$_2$ unit cell on the Si(110) surface. The Si ($\upalpha$-FeSi$_2$) directions/planes are given in black/grey (red/light red). Fe atoms are depicted in blue, Si atoms in green. 
}
\label{fig:surface}
\end{figure}

For the growth of $\upalpha$-FeSi$_2$ on Si($111$) \cite{kalt_alpha_islands,berbezier_TEM,kataoka_phase} and Si($001$) \cite{chen_alpha_wires,won_alpha_2006} the commonly reported epitaxial relation is  Si\{$111$\}$\vert\vert \alpha$-FeSi$_2$\{$112$\}.
In this configuration, the lattice mismatch is minimized if Si$ \langle \bar{1}10 \rangle \vert\vert \alpha$-FeSi$_2 \langle 1\bar{1}0 \rangle$ \cite{berbezier_TEM}.
Translated on the Si($110$) surface, this leads to: Si($111$)$\vert\vert \alpha$-FeSi$_2$($112$) and Si[$\bar{1}10$]$\vert\vert \alpha$-FeSi$_2$[$1\bar{1}0$].
In Fig.\,\,\ref{fig:surface} the corresponding orientation of the $\upalpha$-FeSi$_2$ unit cell on the Si($110$) surface is depicted.
The lattice mismatch (defined as $(a_{Si}-a_{FeSi_2})/a_{Si}$)) is 0.5\,\% along Si[$\bar{1}10$] and 1\,\% along Si[$001$] 
and the tilt angle between $\upalpha$-FeSi$_2$[$110$] and Si[$001$] amounts to 18$^\circ$. 
Furthermore, this configuration implies that  $\upalpha$-FeSi$_2$[$44\bar{1}$] is 0.5$^\circ$ off Si[$001$].

\begin{figure}[t]
 \centering
 \includegraphics[width=0.98\columnwidth]{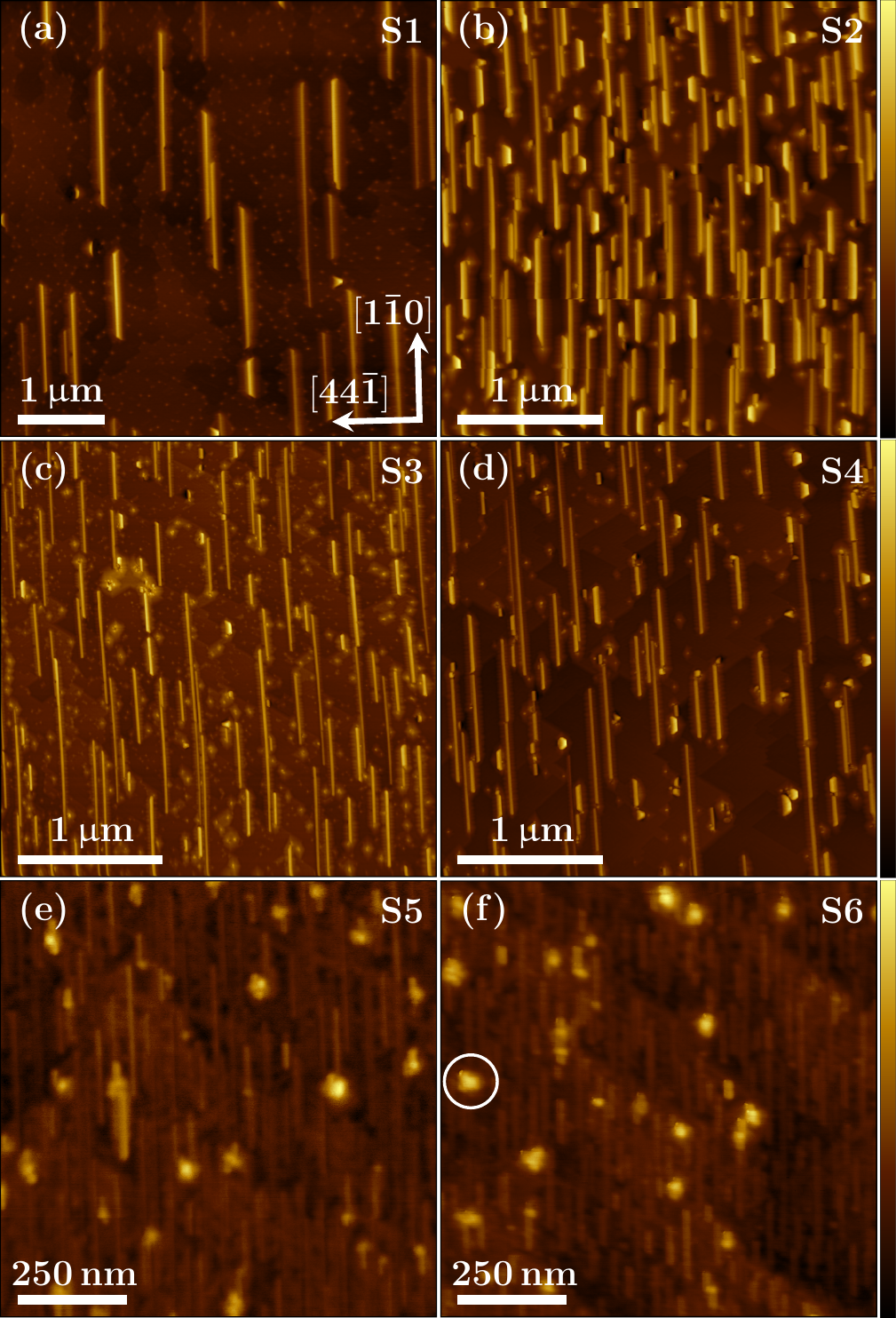}
 \caption{
AFM images of the indicated samples with (a) height scale (hs): 0\,-\,59\,nm, (b) hs: 0\,-\,39\,nm, (c) hs: 0\,-\,30\,nm, (d) hs: 0\,-\,37\,nm, (e) hs: 0\,-\,10\,nm, and (f) hs: 0\,-\,12\,nm. In (a) the crystallographic directions of the $\upalpha$-FeSi$_2$ crystal are indicated. The white circle in (f) marks an exemplary copper contamination. 
}
 \label{fig:AFM}
\end{figure}

Figure \ref{fig:AFM} shows an overview of the AFM images of S1-S6.
For all samples the NWs are unidirectionally aligned along Si[$\bar{1}10$], as reported for the growth of FeSi$_2$ on Si($110$) \cite{endotaxial_nw_prl,tetra,cubic3,cubic1}.
The epitaxial relation discussed above implies that the NWs are formed along Si[$\bar{1}10$]$ \vert \vert \upalpha$-FeSi$_2$[$1\bar{1}0$].
Furthermore, due to the small deviation of 0.5$^\circ$ we approximate that Si[$001$]$\vert \vert \upalpha$-FeSi$_2$[$44\bar{1}$]  (Fig.\,\,\ref{fig:AFM} (a)). 
The average width $\bar{w}$ of the NWs, calculated from AFM line scans \cite{sm}, are given in Table \ref{tab:samples}.
As expected, an increase of the growth temperature $T_G$ or the amount of deposited iron $\theta_{Fe}$ leads to NWs with larger dimensions.
The AFM images of S5 and S6 exhibit additional round islands, an example is marked in Fig.\,\,\ref{fig:AFM}\,(f).
These structures occur after the removal of the native SiO$_2$ layer and by x-ray photoelectron spectroscopy were identified as a copper contamination.
Since the NIS technique is sensitive solely to the $^{57}$Fe nuclei \cite{Seto_PRL_NIS,Sturhahn_PRL_NIS,NIS_selctivity}, the Cu islands do not contribute to the obtained PDOS of the NWs.

\subsection{Lattice dynamics}

\begin{figure}[t!]
 \centering
 \includegraphics[width=0.99\columnwidth]{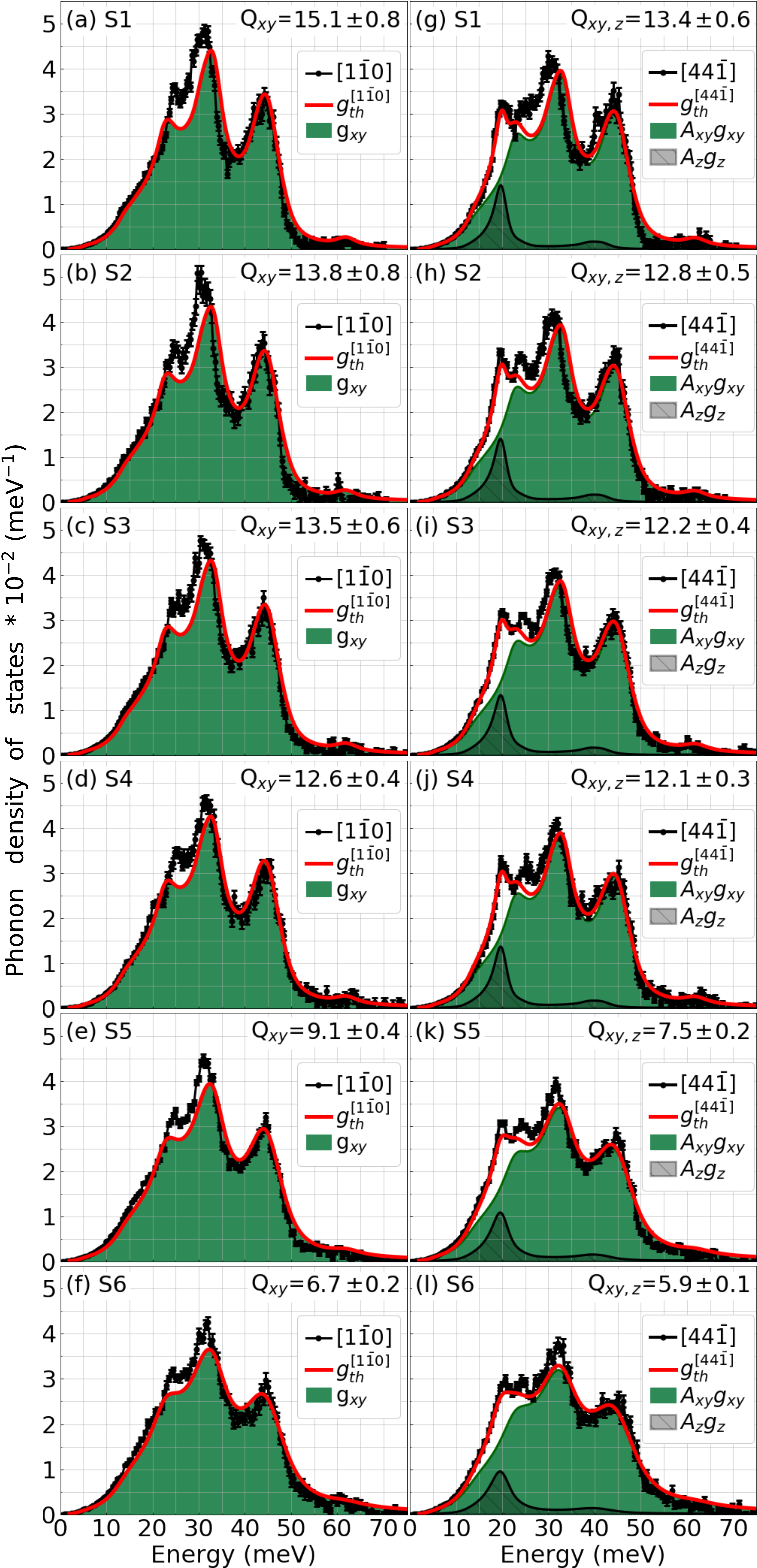}
 \caption{
Fe-partial PDOS of the indicated samples measured (a)\,-\,(f) along $\upalpha$-FeSi$_2$[$1\bar{1}0$] and (g)\,-\,(l) along $\upalpha$-FeSi$_2$[$44\bar{1}$].
The given error bars represent the 1-$\upsigma$ uncertainty.
The experimental data (symbols) is  compared with the results of the least squares fit (solid red line), decomposed into its weighted \textit{xy}\,\,($A_{xy} \, g_{xy}$) and \textit{z}\,\,($A_{z} \, g_{z}$) contributions (for details see text).
The resulting quality factors $Q_{xy}$ and $Q_{xy,z}$ are also given. 
}
 \label{fig:PDOS}
\end{figure}

In Fig.\,\,\ref{fig:PDOS} the Fe-partial PDOS of S1-S6 obtained with the wave vector of the x-ray beam being parallel to $\upalpha$-FeSi$_2$[$1\bar{1}0$] (left column) and $\upalpha$-FeSi$_2$[$44\bar{1}$] (right column) are depicted.
A comparison of the PDOS obtained along and across the NWs shows a vibrational anisotropy with pronounced differences around 20\,meV.
Furthermore, the reduction of the average NW width $\bar{w}$  from 24\,nm (S1) to 3\,nm (S6) leads to a broadening of the peaks.

Previous \textit{ab initio} calculations of the direction-projected Fe-partial PDOS of the tetragonal $\upalpha$-FeSi$_2$ showed a decoupling of vibrations with \textit{xy}- and \textit{z}-polarization \cite{kalt_alpha_islands}.
The Fe-partial PDOS of the \textit{xy}-polarized vibrations consists of peaks at 24, 33, and 45\,meV, while the \textit{z}-polarized vibrations are mostly localized at 20\,meV  with a minor plateau around 40\,meV.
The experimental PDOS obtained with the wave vector of the x-ray beam being parallel to a certain crystallographic direction of the NWs is composed of a specific combination of \textit{xy}- and \textit{z}-polarized phonons \cite{kalt_alpha_islands,chumakov_anisotropic_nis_FeBO3,kohn_anisotropic_nis_theo}.
The relative contributions of \textit{xy}-, ($A_{xy}$) and \textit{z}- ($A_{z}$) polarized phonons can be calculated \cite{sm} considering the orientation of the $\upalpha$-FeSi$_2$ unit cell and amount to $A^{[1\bar{1}0]}_{xy}$=1 and $A^{[1\bar{1}0]}_{z}$=0 for $\upalpha$-FeSi$_2$[$1\bar{1}0$] and $A^{[44\bar{1}]}_{xy}$=0.9 and $A^{[44\bar{1}]}_{z}$=0.1 for $\upalpha$-FeSi$_2$[$44\bar{1}$]. 
Consequently, the observed vibrational anisotropy originates from the specific orientation of the $\upalpha$-FeSi$_2$ unit cell on the Si(110) surface (see Fig.\,3).

The observed damping of the PDOS features upon reduction of the NW size can be quantified by comparison of the experimental results with the \textit{ab initio} calculations \cite{kalt_alpha_islands}. 
The damping originates  from  phonon scattering at defects at interfaces and surfaces, as well as within the crystal \cite{fultz_dho} and  can be described by the damped harmonic oscillator (DHO) function \cite{Faak1}.
The DHO function introduces an energy-dependent broadening of the spectral features quantified by the quality factor $Q$, which is inversely proportional to the strength of the damping.
The experimental PDOS data were modeled by convolution of the \textit{ab initio} calculated  PDOS, obtained for a 0.5\,\% tensile strained $\upalpha$-FeSi$_2$ crystal, with the DHO function.
The strength of the damping in the respective sample is quantified by the $Q$ values obtained using  the  least-squares method.

For measurements along  $\upalpha$-FeSi$_2$[$1\bar{1}0$] (along the NWs), the PDOS consists of \textit{xy}-polarized vibrations only and the experimental data can be described by:
\begin{equation}
g^{[1\bar{1}0]}_{th}\,=g_{xy}(E,Q_{xy}),
\end{equation} 
with $g_{xy}$ being the  \textit{ab initio} calculated \textit{xy}-polarized Fe-partial PDOS convoluted with  the DHO function with a quality factor $Q_{xy}$.
In Fig. \ref{fig:PDOS} (a)-(f)  $g^{[1\bar{1}0]}_{th}$ is compared with the respective experimental PDOS.
In general, a very good agreement is observed between experiment and theory.
While the peak around 45\,meV occurs at the same positions in the experimentally determined and \textit{ab initio} calculated PDOS, the minor peak around 25\,meV is shifted by 1.5\,meV to lower energy  and the peak around 33\,meV is shifted by about 1\,meV to higher energy in the \textit{ab initio} calculated PDOS.
Most likely these differences occur due to a more complicated strain distribution in the $\upalpha$-FeSi$_2$ crystal than the assumed isotropic 0.5\,\% tensile strain.

\begin{figure}[t!]
\includegraphics[width=0.99\columnwidth]{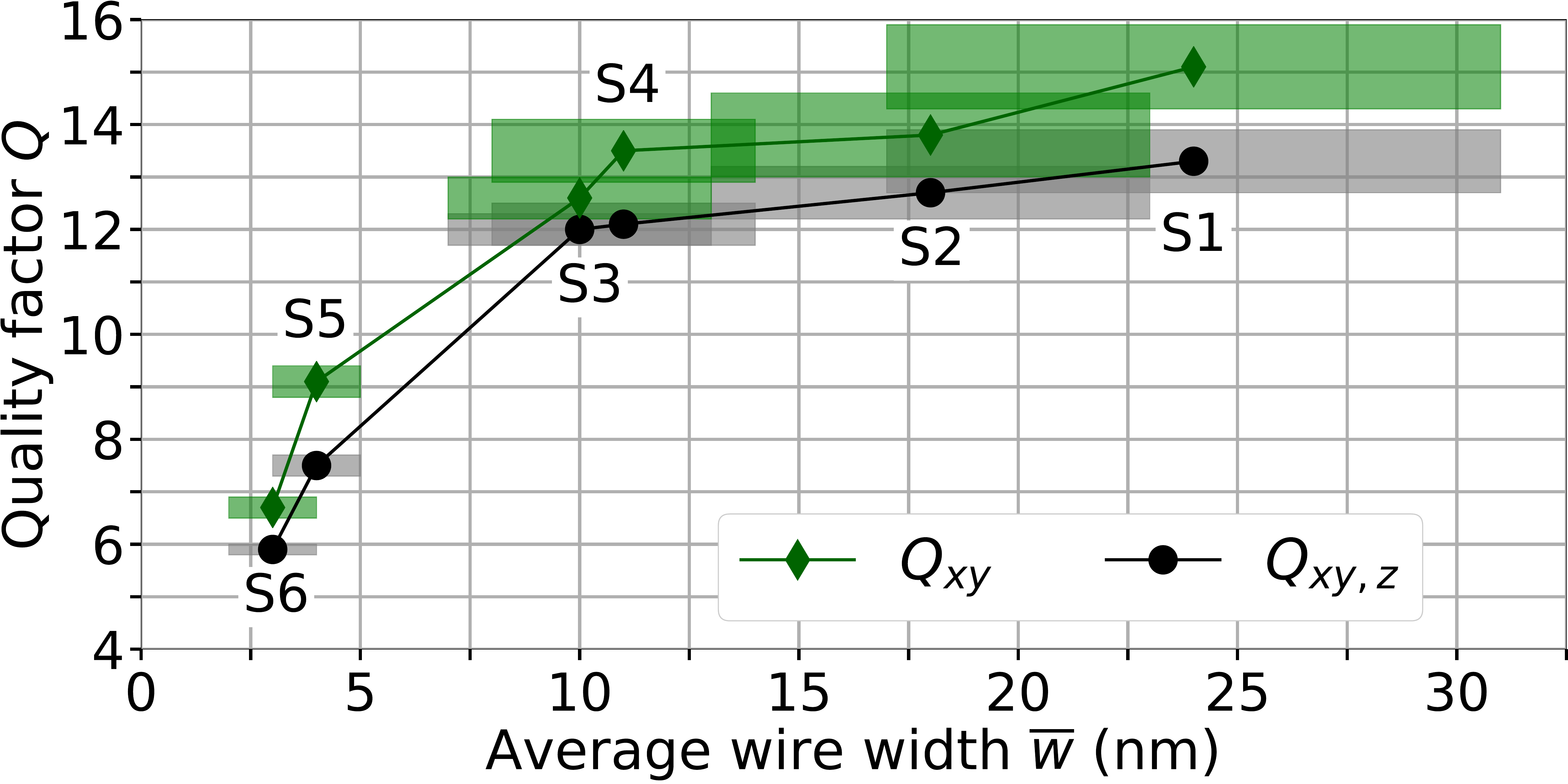}
\caption{Quality factors $Q_{xy}$  and $Q_{xy,z}$ (Fig. \ref{fig:PDOS}) as a function of average NW width $\bar{w}$ (Table \ref{tab:samples}). The shaded boxes denote the uncertainties  in $Q_{xy}$, $Q_{xy,z}$ and  $\bar{w}$. 
}
\label{fig:Q_values}
\end{figure}

The PDOS obtained along $\upalpha$-FeSi$_2$[$44\bar{1}$] (across the NWs) is  modeled by the weighted sum of the \textit{ab initio} calculated \textit{xy}- and \textit{z}-polarized PDOS, convoluted with the DHO function with a quality factor $Q_{xy,z}$:\newline
\begin{equation}
g^{[ 44\bar{1} ]}_{th}=  A^{[44\bar{1}]}_{xy} \cdot g_{xy}(E,Q_{xy,z}) +  A^{[44\bar{1}]}_{z}\cdot g_{z}(E,Q_{xy,z}).
\end{equation} 
In Fig. \ref{fig:PDOS} (g)-(l) $g^{[44\bar{1}]}_{th}$ is plotted with the respective \textit{xy}- ($A_{xy}g_{xy}$) and \textit{z}- ($A_{z}g_{z}$) contributions and the corresponding $Q_{xy,z}$ values obtained from the fit \cite{note_Axy_Az}.
A very good agreement  between experiment and theory is observed.
The peak of the \textit{z}-polarized phonons around 20\,meV occurs at the same energies in theory and experiment. 
The \textit{xy}-polarized vibrations along $\upalpha$-FeSi$_2$[$44\bar{1}$] are also well reproduced by the model, except for small shifts of the peaks at 25 and 33\,meV, which are also observed in the measurements along $\upalpha$-FeSi$_2$[$1\bar{1}0$] and are attributed to a complex strain distribution.
On average, the $Q_{xy,z}$ values obtained for the PDOS across the NWs are reduced by 10\,\% compared to the $Q_{xy}$ values obtained for the PDOS along the NWs. 
The reason for the slightly stronger damping of the phonons propagating across the NWs could be the smaller size of the  $\upalpha$-FeSi$_2$ crystal along this direction.

In Fig.\,\,\ref{fig:Q_values} the quality factors obtained from the least squares fits for S1-S6 along [$1\bar{1}0$] and [$44\bar{1}$] are depicted as a function of $\bar{w}$. 
The $Q_{xy}$ and $Q_{xy,z}$ values of S1-S4 show a slight decrease in the range of 24\,nm\,$\geq \bar{w}\geq$\,10\,nm, whereas upon reduction of $\bar{w}$ below 10\,nm in S5 and S6 $Q_{xy}$ and $Q_{xy,z}$ are significantly reduced.
To comprehend this trend, the interface-to-volume ratio of the NWs has to be considered. 
In the volume part, i.e. the core of the NWs, the atoms are located in a bulk-like environment with a high degree of crystalline order.
At the interface towards the substrate the defect density is generally increased  and thus the scattering of phonons is enhanced.
For the small wires  of S5 and S6 the interface-to-volume ratio is significantly higher and consequently $Q_{xy}$ and $Q_{xy,z}$ are distinctly reduced.

 \begin{figure}[t!]
 \centering
 \includegraphics[width=0.98\columnwidth]{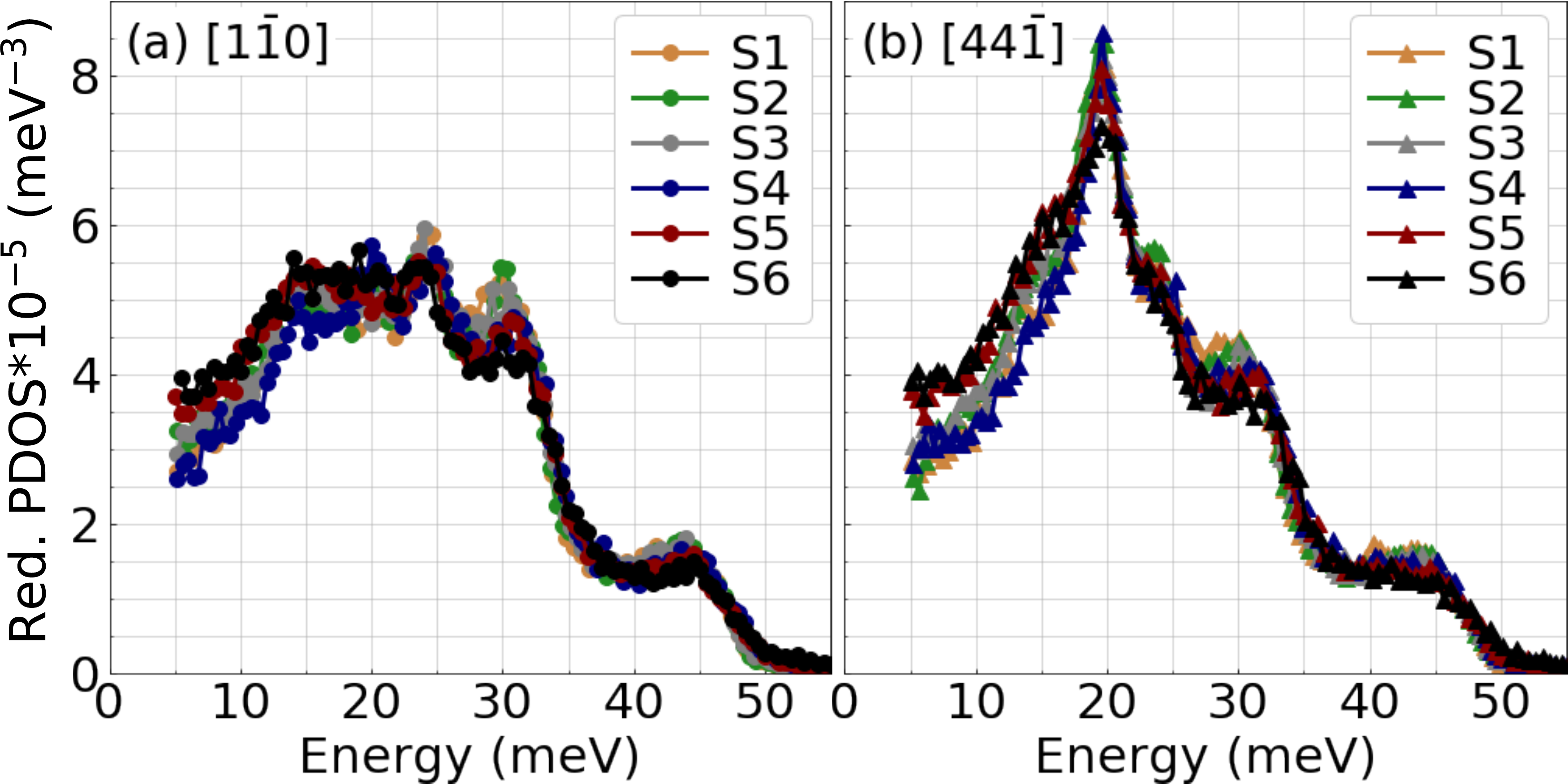}
 \caption{
Fe-partial reduced PDOS [g(E)/E$^2$] of the indicated samples projected (a) along  $\upalpha$-FeSi$_2$[$1\bar{1}0$] and (b) along $\upalpha$-FeSi$_2$[$44\bar{1}$].
}
\label{fig:reddos}
\end{figure}

\begin{table*}[t]
\small
 \caption{Fe-partial mean force constant $F$, mean square displacement $\langle x^2 \rangle$, vibrational entropy $S_V$ and heat capacity $C_V$ calculated from the \textit{ab initio} and  the experimental PDOS of S1-S6 projected along $\upalpha$-FeSi$_2$[$1\bar{1}0$] and $\upalpha$-FeSi$_2$[$44\bar{1}$]. The coefficient $ \alpha$ derived from the low-energy part of the reduced PDOS [$g(E)/E^2\,=\,\alpha$]  and the sound velocity $v_S$ are also given.
}  
\label{tab:TDP_sound}
  \begin{tabular*}{\textwidth}{@{\extracolsep{\fill}}cccccccc}
    \hline
       \hline\\[-3mm]
  		& direction      	&  $F\,(N/m)$		& $\langle x^2 \rangle$\,(\AA$^2$)		& $S_V$\,($k_B/atom$) 			& $C_V$\,($k_B/atom$) 			&$\alpha$	($10^{-5}meV^{-3}$)	& $v_S$    $(m/s)$						\\ 
\hline\\[-3mm]
Theory  &	[$1\bar{1}0$]	&	254\,$\pm$5		&	0.0096\,$\pm$\,0.0002	  			&		2.62\,$\pm$0.02			&		2.60\,$\pm$0.02			& 		-					    & 		4988							\\ 
  		&	[$44\bar{1}$]	&	245\,$\pm$5		&	0.0096\,$\pm$\,0.0002			  	&		 2.68\,$\pm$\,0.02	 	&		2.61\,$\pm$\,0.02 		& 		-					    & 		-								\\    
S1  	&	[$1\bar{1}0$]	&	249\,$\pm$5		&	0.0100\,$\pm$\,0.0002	  			&		2.69\,$\pm$0.02			&		2.61\,$\pm$0.02			& 		3.20$\pm$0.04  			& 		4700 $\pm$\,\,\, 50				\\ 
  		&	[$44\bar{1}$]	&	245\,$\pm$5		&	0.0104\,$\pm$\,0.0002		  		&		2.74\,$\pm$0.02			&		2.62\,$\pm$0.02			& 		3.03$\pm$0.07   		& 		4780 $\pm$ 100					\\  
S2  	&	[$1\bar{1}0$]	&	250\,$\pm$5		&	0.0101\,$\pm$\,0.0002		  		&		2.67\,$\pm$0.02			&		2.60\,$\pm$0.02     	& 		3.51$\pm$0.07  			& 		4550 $\pm$ 100					\\ 		
  		&	[$44\bar{1}$]	&	245\,$\pm$5		&	0.0105\,$\pm$\,0.0002				&		2.73\,$\pm$0.02			&		2.61\,$\pm$0.02			& 		3.24$\pm$0.08   		& 		4670 $\pm$ 120					\\    
S3 		&	[$1\bar{1}0$]	&	248\,$\pm$5		&	0.0101\,$\pm$\,0.0002		  		&		2.67\,$\pm$0.02			&		2.60\,$\pm$0.02			& 		3.38$\pm$0.06   		& 		4600 $\pm$\,\,\, 80				\\  
  		&	[$44\bar{1}$]	&	247\,$\pm$5		&	0.0105\,$\pm$\,0.0002				&		2.72\,$\pm$0.02			&		2.61\,$\pm$0.02			& 		3.37$\pm$0.07  			& 		4610 $\pm$ 100					\\   
S4  	&	[$1\bar{1}0$]	&	256\,$\pm$5		&	0.0096\,$\pm$\,0.0002		  		&		2.64\,$\pm$0.02			&		2.59\,$\pm$0.02			& 		3.01$\pm$0.09   		& 		4780 $\pm$ 140					\\
  		&	[$44\bar{1}$]	&	251\,$\pm$5		&	0.0102\,$\pm$\,0.0002	  			&		2.69\,$\pm$0.02			&		2.59\,$\pm$0.02			& 		3.13$\pm$0.04   		& 		4730 $\pm$\,\,\, 60				\\    
S5  	&	[$1\bar{1}0$]	&	259\,$\pm$5		&	0.0104\,$\pm$\,0.0002		  		&		2.66\,$\pm$0.02			&		2.59\,$\pm$0.02			& 		3.69$\pm$0.04   		& 		4480 $\pm$\,\,\, 50				\\ 
  		&	[$44\bar{1}$]	&	250\,$\pm$5		&	0.0110\,$\pm$\,0.0002				&		2.74\,$\pm$0.02			&		2.60\,$\pm$0.02			& 		3.82$\pm$0.05   		& 		4420 $\pm$\,\,\, 60				\\ 
S6  	&	[$1\bar{1}0$]	&	255\,$\pm$5		&	0.0105\,$\pm$\,0.0002		  		&		2.68\,$\pm$0.02			&		2.59\,$\pm$0.02			& 		3.96$\pm$0.05   		& 		4370 $\pm$\,\,\, 60				\\ 
  		&	[$44\bar{1}$]	&	255\,$\pm$5		&	0.0110\,$\pm$\,0.0002				&		2.71\,$\pm$0.02			&		2.60\,$\pm$0.02 		& 		4.00$\pm$0.05   		& 		4360 $\pm$\,\,\, 60				\\ 
        \hline
           \hline
  \end{tabular*}
\end{table*}

A study of the lattice dynamics of epitaxial $\upalpha$-FeSi$_2$ nanoislands on Si(111) unveiled a polarization-dependence of the phonon damping, i.e. a stronger damping of \textit{z}-polarized phonons, in islands with average heights below 10\,nm \cite{kalt_alpha_islands}.
To examine if this effect is also present in the investigated NWs, the PDOS obtained along $\upalpha$-FeSi$_2$[$44\bar{1}$] were fitted by the weighted sum of the \textit{ab initio} calculated \textit{xy}- and \textit{z}-polarized PDOS convoluted with  DHO functions with independent quality factors.
Although the results indicate that this effect might also be present in S5 and S6, the low intensity of the peak at 20\,meV does not allow for a definite conclusion.

In Fig. \ref{fig:reddos} the Fe-partial reduced PDOS [g(E)/E$^2$] of S1-S6 is shown.
Along both directions no systematic increase in low-energy states is observed with reduction of $\bar{w}$ from 24\,nm (S1) to 10\,nm (S3).
However, the samples with the smallest NWs, S5 and S6, show an increase of states in the region from 5 to 15\,meV.
Such an enhancement of low-energy states has been observed in the PDOS of thin films \cite{pradip_EuO_nscale,Fe3Si_interface, slezak1} and surfaces \cite{stankov1}.
It is attributed to interface/surface-specific vibrational modes, which are more pronounced in the smallest NWs.

A comparison of the PDOS and reduced PDOS of the capped and uncapped NWs of S3 and S4 unveils only minor deviations (see figure in supplementary material \cite{sm}).
The negligible influence of the capping layer can be explained by the endotaxial growth of the NWs, which results in a large NW/substrate interface and a small fraction of atoms located at the surface of the NWs.

\subsection{Thermodynamic and elastic properties} 

The thermodynamic and elastic properties obtained from the \textit{ab initio} calculated and experimentally determined PDOS \cite{NIS_selctivity} are given in Table \ref{tab:TDP_sound}.
The experiments show an average decrease of the mean force constant $F$ by  1.6\,\% along  [$44\bar{1}$]  compared to [$1\bar{1}0$], while the mean square displacement $\langle x^2 \rangle$ and the vibrational entropy $S_V$ on average increase by 5\,\% and 2\,\%, respectively.
These differences originate from the vibrational anisotropy of the tetragonal $\upalpha$-FeSi$_2$  unit cell. 
The contribution of \textit{z}-polarized phonons along $\upalpha$-FeSi$_2$[$44\bar{1}$] induces the observed softening of the crystal compared to $\upalpha$-FeSi$_2$[$1\bar{1}0$].

The reduction of $\bar{w}$ of the NWs from S1-S6 leads to an increase of $\langle x^2 \rangle$ by 5\,\% along [$1\bar{1}0$] and by 5.8\,\% along [$44\bar{1}$].
This is in agreement with the trend observed in the $\sigma^2$ values obtained for the Fe-Si scattering path by modeling the EXAFS data (see Table. \ref{tab:tab_EXAFS}).
$F$ is also slightly increased from S1 to S6, most likely due to the enhancement of high energy states above the cutoff energy, induced by the broadening of the peak at 45\,meV.
A similar behavior  was observed in $\upalpha$-FeSi$_2$ nanoislands \cite{kalt_alpha_islands}.
A comparison with the theoretically expected values for bulk $\upalpha$-FeSi$_2$ shows that the $\langle x^2 \rangle$ and $S_V$ values are on average increased by 8\,\% and  2\,\%, respectively, in the smallest NWs.
The values of the heat capacity $C_V$ coincide within the uncertainty for S1-S6 in both directions.

The low-energy part of the PDOS in all samples can be described by the Debye model: $g(E)\,=\,\alpha E^2$.
The coefficient $\alpha$  is on average increasing as the NWs size is decreasing (Fig.\,\,\ref{fig:reddos}).
Using $\alpha$, the sound velocity $v_S$ of S1-S6 was calculated \cite{hu_sos}. 
For comparison the theoretical value for [$1\bar{1}0$], determined from the slopes of three acoustic branches calculated along {\sffamily $\Upgamma$}-{\sffamily M} direction, is also given. 
The experimental values are clearly reduced compared to the theoretical value. 
Reason for these differences is that a perfect crystal is assumed for the \textit{ab initio} calculations, whereas in the NWs the propagation of sound waves is decelerated by scattering at defects, which are mostly present at interfaces.
Since the interface-to-volume ratio is increased when the NWs dimensions are reduced, $v_S$ is also reduced from S1-S6 by 9\,\%.

\section{Conclusions}\label{Conclusions}

Endotaxial FeSi$_2$ NWs were grown on Si(110) by reactive deposition epitaxy.
Systematic RHEED and AFM studies unveiled the formation of single-crystalline, unidirectionally aligned NWs with average widths $\bar{w}$ from 24 to 3\,nm and lengths from several $\upmu$m to about 100\,nm.
A combined experimental and theoretical EXAFS study demonstrated that the NWs exhibit the metastable, surface-stabilized $\upalpha$-FeSi$_2$ crystal structure.

The Fe-partial PDOS was determined along and across the NWs by NIS  experiments performed at room temperature.
A pronounced vibrational anisotropy originating from the specific orientation of the tetragonal $\upalpha$-FeSi$_2$ unit cell on the Si(110) surface was unveiled.
Modeling of the experimental data with first-principles calculations showed that upon reduction of $\bar{w}$ from 24 to 3\,nm the features of PDOS broaden significantly.
This is attributed to phonon scattering at the NW/substrate interface, which is particularly strong in the smallest NWs characterized with the highest interface-to-volume ratio.
Furthermore, the reduction of $\bar{w}$ from 24\,nm to 3\,nm leads to  an increase of the mean square displacement by 5\,\% and a reduction of the sound velocity by 9\,\%.
The damping of lattice vibrations is slightly stronger across the nanowires, due to the smaller size of the  $\upalpha$-FeSi$_2$ crystal along this direction.
A comparison of the PDOS of NWs with identical sizes  measured with and without a capping layer demonstrates that the influence of surface-specific vibrational modes is negligible due to the endotaxial character of the NWs.\\
The presented results on the lattice dynamics and thermoelastic properties of FeSi$_2$ nanowires are expected to be generally valid for the technologically important class of endotaxial silicide nanowires.

\begin{acknowledgments}
S.S. acknowledges the financial support by  the  Helmholtz  Association  (VH-NG-625)  and  BMBF (05K16VK4).  P.P. acknowledges support by the Narodowe Centrum Nauki (NCN, National Science Centre)
under Project No. 2017/25/B/ST3/02586 and the access to ESRF financed by the Polish Ministry of Science and High Education, decision number: DIR/WK/2016/19. The European Synchrotron Radiation Facility is acknowledged for beamtime provision at the Nuclear Resonance beamline ID18. We thank Mr. J.-P. Celse for technical assistance during the experiment at ID18. We acknowledge DESY (Hamburg, Germany), a member of the Helmholtz Association HGF, for the provision of experimental facilities. Parts of this research were carried out at the High Resolution Dynamics Beamline P01 at PETRA III. We thank Mr. C. Hagemeister and Mr. F.-U. Dill  for technical assistance during the experiment at P01.
\end{acknowledgments}


\begin{thebibliography}{0}
\expandafter\ifx\csname natexlab\endcsname\relax\def\natexlab#1{#1}\fi
\expandafter\ifx\csname bibnamefont\endcsname\relax
  \def\bibnamefont#1{#1}\fi
\expandafter\ifx\csname bibfnamefont\endcsname\relax
  \def\bibfnamefont#1{#1}\fi
\expandafter\ifx\csname citenamefont\endcsname\relax
  \def\citenamefont#1{#1}\fi
\expandafter\ifx\csname url\endcsname\relax
  \def\url#1{\texttt{#1}}\fi
\expandafter\ifx\csname urlprefix\endcsname\relax\def\urlprefix{URL }\fi
\providecommand{\bibinfo}[2]{#2}
\providecommand{\eprint}[2][]{\url{#2}}

\end{thebibliography}


\begin{thebibliography}{1}
\expandafter\ifx\csname natexlab\endcsname\relax\def\natexlab#1{#1}\fi
\expandafter\ifx\csname bibnamefont\endcsname\relax
  \def\bibnamefont#1{#1}\fi
\expandafter\ifx\csname bibfnamefont\endcsname\relax
  \def\bibfnamefont#1{#1}\fi
\expandafter\ifx\csname citenamefont\endcsname\relax
  \def\citenamefont#1{#1}\fi
\expandafter\ifx\csname url\endcsname\relax
  \def\url#1{\texttt{#1}}\fi
\expandafter\ifx\csname urlprefix\endcsname\relax\def\urlprefix{URL }\fi
\providecommand{\bibinfo}[2]{#2}
\providecommand{\eprint}[2][]{\url{#2}}



\bibitem[{\citenamefont{Murarka}(1995)\citenamefont{}}]{Murarka_silicides_microelectronics}
    \bibinfo{author}{\bibfnamefont{S.P.}~\bibnamefont{Murarka}}, 
    \textit{\bibinfo{journal}{Intermetallics}} \textbf{\bibinfo{volume}{3}}, \bibinfo{pages}{173}  {\bibinfo{year}{(1995)}}.

\bibitem[{\citenamefont{Chen_book}(2004)\citenamefont{}}]{Chen_book}
\bibinfo{author}{\bibfnamefont{L.J.}~\bibnamefont{Chen}},
     \textit{\bibinfo{journal}{Silicide technology for integrated circuits}} {\bibinfo{year}{Institution of Electrical Engineers}}, \bibinfo{pages}{London  (2004)}.

\bibitem[{\citenamefont{Chen}(2005)\citenamefont{}}]{chen_silicides_microelectronics}
    \bibinfo{author}{\bibfnamefont{L.J.}~\bibnamefont{Chen}}, 
    \textit{\bibinfo{journal}{JOM}} \textbf{\bibinfo{volume}{57}}, \bibinfo{pages}{24}  {\bibinfo{year}{(2005)}}.

\bibitem[{\citenamefont{Bennett et~al.}(2011)\citenamefont{}}]{endotaxial_review}
\bibinfo{author}{\bibfnamefont{P.A.}~\bibnamefont{Bennett}},
\bibinfo{author}{\bibfnamefont{Z.}~\bibnamefont{He}},
  \bibinfo{author}{\bibfnamefont{D.J.}~\bibnamefont{Smith}}, \bibnamefont{and}
  \bibinfo{author}{\bibfnamefont{F.M.}~\bibnamefont{Ross}},
     \textit{\bibinfo{journal}{Thin Solid Films}} \textbf{\bibinfo{volume}{519}}, \bibinfo{pages}{8434} {\bibinfo{year}{(2011)}}.

\bibitem[{\citenamefont{He et~al.}(2004)\citenamefont{}}]{endotaxial_nw_prl}
\bibinfo{author}{\bibfnamefont{Z.}~\bibnamefont{He}},
  \bibinfo{author}{\bibfnamefont{D.J.}~\bibnamefont{Smith}}, \bibnamefont{and}
  \bibinfo{author}{\bibfnamefont{P.A.}~\bibnamefont{Bennett}},
     \textit{\bibinfo{journal}{Phys. Rev. Lett.}} \textbf{\bibinfo{volume}{93}}, \bibinfo{pages}{256102} {\bibinfo{year}{(2004)}}.

\bibitem[{\citenamefont{Mahato et~al.}(2017)\citenamefont{}}]{nanotechnology}
\bibinfo{author}{\bibfnamefont{J.C.}~\bibnamefont{Mahato}},
\bibinfo{author}{\bibfnamefont{D.}~\bibnamefont{Das}},
  \bibinfo{author}{\bibfnamefont{N.}~\bibnamefont{Banu}},
  \bibinfo{author}{\bibfnamefont{B.}~\bibnamefont{Satpati}}, \bibnamefont{and}
  \bibinfo{author}{\bibfnamefont{B.N.}~\bibnamefont{Dev}},
     \textit{\bibinfo{journal}{Nanotechnology}} \textbf{\bibinfo{volume}{28}}, \bibinfo{pages}{425603}  {\bibinfo{year}{(2017)}}.

\bibitem[{\citenamefont{Liang et~al.}(2006)\citenamefont{}}]{cubic1}
\bibinfo{author}{\bibfnamefont{S.}~\bibnamefont{Liang}},
\bibinfo{author}{\bibfnamefont{R.}~\bibnamefont{Islam}},
  \bibinfo{author}{\bibfnamefont{D.J.}~\bibnamefont{Smith}},
  \bibinfo{author}{\bibfnamefont{P.A.}~\bibnamefont{Bennett}},
  \bibinfo{author}{\bibfnamefont{J.R.}~\bibnamefont{O'Brien}}, \bibnamefont{and}
  \bibinfo{author}{\bibfnamefont{B.}~\bibnamefont{Taylor}},
     \textit{\bibinfo{journal}{Appl. Phys. Lett.}} \textbf{\bibinfo{volume}{88}}, \bibinfo{pages}{113111}  {\bibinfo{year}{(2006)}}.

\bibitem[{\citenamefont{Liang et~al.}(2006)\citenamefont{}}]{cubic2}
\bibinfo{author}{\bibfnamefont{S.}~\bibnamefont{Liang}},
\bibinfo{author}{\bibfnamefont{R.}~\bibnamefont{Islam}},
  \bibinfo{author}{\bibfnamefont{D.J.}~\bibnamefont{Smith}}, \bibnamefont{and}
  \bibinfo{author}{\bibfnamefont{P.A.}~\bibnamefont{Bennett}},
     \textit{\bibinfo{journal}{J. Cryst. Growth}} \textbf{\bibinfo{volume}{295}}, \bibinfo{pages}{166} {\bibinfo{year}{(2006)}}.

\bibitem[{\citenamefont{Das et~al.}(2014)\citenamefont{}}]{cubic3}
\bibinfo{author}{\bibfnamefont{D.}~\bibnamefont{Das}},
\bibinfo{author}{\bibfnamefont{J.C.}~\bibnamefont{Mahato}},
\bibinfo{author}{\bibfnamefont{B.}~\bibnamefont{Bisi}},
  \bibinfo{author}{\bibfnamefont{B.}~\bibnamefont{Satpati}}, \bibnamefont{and}
  \bibinfo{author}{\bibfnamefont{B.N.}~\bibnamefont{Dev}},
     \textit{\bibinfo{journal}{Appl. Phys. Lett.}} \textbf{\bibinfo{volume}{105}}, \bibinfo{pages}{191606} {\bibinfo{year}{(2014)}}.

\bibitem[{\citenamefont{Zou et~al.}(2017)\citenamefont{}}]{tetra}
\bibinfo{author}{\bibfnamefont{Z.-Q.}~\bibnamefont{Zou}},
\bibinfo{author}{\bibfnamefont{X.}~\bibnamefont{Li}},
\bibinfo{author}{\bibfnamefont{X.-Y.}~\bibnamefont{Liu}},
  \bibinfo{author}{\bibfnamefont{K.-J.}~\bibnamefont{Shi}}, \bibnamefont{and}
  \bibinfo{author}{\bibfnamefont{X.-Q.}~\bibnamefont{Guo}},
     \textit{\bibinfo{journal}{Appl. Surf. Sci.}} \textbf{\bibinfo{volume}{399}}, \bibinfo{pages}{200} {\bibinfo{year}{(2017)}}.

\bibitem[{\citenamefont{Bozyigit et~al.}(2016)\citenamefont{}}]{Bozyigit}
\bibinfo{author}{\bibfnamefont{D.}~\bibnamefont{Bozyigit}},
\bibinfo{author}{\bibfnamefont{N.}~\bibnamefont{Yazdani}},
\bibinfo{author}{\bibfnamefont{M.}~\bibnamefont{Yarema}},
\bibinfo{author}{\bibfnamefont{O.}~\bibnamefont{Yarema}},
\bibinfo{author}{\bibfnamefont{W.M.M.}~\bibnamefont{Li}},
\bibinfo{author}{\bibfnamefont{S.}~\bibnamefont{Volk}},
\bibinfo{author}{\bibfnamefont{K.}~\bibnamefont{Vuttivorakulchai}},
\bibinfo{author}{\bibfnamefont{M.}~\bibnamefont{Luisier}},
\bibinfo{author}{\bibfnamefont{F.}~\bibnamefont{Juranyi}}, \bibnamefont{and}
\bibinfo{author}{\bibfnamefont{V.}~\bibnamefont{Wood}},
     \textit{\bibinfo{journal}{Nature}} \textbf{\bibinfo{volume}{531}}, \bibinfo{pages}{618} {\bibinfo{year}{(2016)}}.

\bibitem[{\citenamefont{Hepplestone}(2005)\citenamefont{}}]{hepplestone_apl_2005}
\bibinfo{author}{\bibfnamefont{S.P.}~\bibnamefont{Hepplestone}}, \bibnamefont{and}
\bibinfo{author}{\bibfnamefont{G.P.}~\bibnamefont{Srivastava}},
     \textit{\bibinfo{journal}{Appl. Phys. Lett.}} \textbf{\bibinfo{volume}{87}}, \bibinfo{pages}{231906} {\bibinfo{year}{(2005)}}.

\bibitem[{\citenamefont{Steinh{\"o}gl et~al.}(2002)\citenamefont{}}]{nanowires_resis}
\bibinfo{author}{\bibfnamefont{W.}~\bibnamefont{Steinh{\"o}gl}},
\bibinfo{author}{\bibfnamefont{G.}~\bibnamefont{Schindler}},
  \bibinfo{author}{\bibfnamefont{G.}~\bibnamefont{Steinlesberger}}, \bibnamefont{and}
  \bibinfo{author}{\bibfnamefont{M.}~\bibnamefont{Engelhardt}},
     \textit{\bibinfo{journal}{Phys. Rev. B}} \textbf{\bibinfo{volume}{66}}, \bibinfo{pages}{075414} {\bibinfo{year}{(2002)}}.

\bibitem[{\citenamefont{Rideau et~al.}(2011)\citenamefont{}}]{nanowires_phonon}
\bibinfo{author}{\bibfnamefont{D.}~\bibnamefont{Rideau}},
\bibinfo{author}{\bibfnamefont{W.}~\bibnamefont{Zhang}},
\bibinfo{author}{\bibfnamefont{Y.M.}~\bibnamefont{Niquet}},
\bibinfo{author}{\bibfnamefont{C.}~\bibnamefont{Delerue}},
  \bibinfo{author}{\bibfnamefont{C.}~\bibnamefont{Tavernier}}, \bibnamefont{and}
  \bibinfo{author}{\bibfnamefont{H.}~\bibnamefont{Jaouen}},
     \textit{\bibinfo{booktitle}{2011 International Conference on Simulation of Semiconductor Processes and Devices}}  \bibinfo{pages}{47}  {\bibinfo{year}{(2011)}}.

\bibitem[{\citenamefont{Tobler et~al.}(2015)\citenamefont{}}]{tobler_resistivity}
\bibinfo{author}{\bibfnamefont{S.K.}~\bibnamefont{Tobler}}, \bibnamefont{and}
\bibinfo{author}{\bibfnamefont{P.A.}~\bibnamefont{Bennett}},
     \textit{\bibinfo{journal}{J. Appl. Phys.}} \textbf{\bibinfo{volume}{118}}, \bibinfo{pages}{125305} {\bibinfo{year}{(2015)}}.

\bibitem[{\citenamefont{Nishiguchi et al.,}(1997)\citenamefont{}}]{theor1}
  \bibinfo{author}{\bibfnamefont{N.}~\bibnamefont{Nishiguchi}},
  \bibinfo{author}{\bibfnamefont{Y.}~\bibnamefont{Ando}}, \bibnamefont{and}
  \bibinfo{author}{\bibfnamefont{M.N.}~\bibnamefont{Wybourne}},
     \textit{\bibinfo{journal}{J. Phys. Condens. Matter}} \textbf{\bibinfo{volume}{9}}, \bibinfo{pages}{5751} {\bibinfo{year}{(1997)}}.

\bibitem[{\citenamefont{Mukdadi et al.,}(1997)\citenamefont{}}]{theor2}
  \bibinfo{author}{\bibfnamefont{O.M.}~\bibnamefont{Mukdadi}},
  \bibinfo{author}{\bibfnamefont{S.K.}~\bibnamefont{Datta}}, \bibnamefont{and}
  \bibinfo{author}{\bibfnamefont{M.L.}~\bibnamefont{Dunn}},
     \textit{\bibinfo{journal}{J. Appl. Phys.}} \textbf{\bibinfo{volume}{97}}, \bibinfo{pages}{074313} {\bibinfo{year}{(2005)}}.

\bibitem[{\citenamefont{Srivastava}(2006)\citenamefont{}}]{theor3}
  \bibinfo{author}{\bibfnamefont{S.P.}~\bibnamefont{Hepplestone}} \bibnamefont{and}
  \bibinfo{author}{\bibfnamefont{G.P.}~\bibnamefont{Srivastava}},
     \textit{\bibinfo{journal}{Nanotechnology}} \textbf{\bibinfo{volume}{17}}, \bibinfo{pages}{3288} {\bibinfo{year}{(2006)}}.

\bibitem[{\citenamefont{Allen}(2007)\citenamefont{}}]{theor4}
  \bibinfo{author}{\bibfnamefont{P.B.}~\bibnamefont{Allen}},
     \textit{\bibinfo{journal}{Nano Lett.}} \textbf{\bibinfo{volume}{7}}, \bibinfo{pages}{11} {\bibinfo{year}{(2007)}}.

\bibitem[{\citenamefont{Mizuno et al.,}(2009)\citenamefont{}}]{theor5}
  \bibinfo{author}{\bibfnamefont{S.}~\bibnamefont{Mizuno}} \bibnamefont{and}
  \bibinfo{author}{\bibfnamefont{N.}~\bibnamefont{Nishiguchi}},
     \textit{\bibinfo{journal}{J. Phys. Condens. Matter}} \textbf{\bibinfo{volume}{21}}, \bibinfo{pages}{195303} {\bibinfo{year}{(2009)}}.

\bibitem[{\citenamefont{Seng\"{u}n}(2010)\citenamefont{}}]{theor7}
  \bibinfo{author}{\bibfnamefont{Y.}~\bibnamefont{S\c{e}ng\"{u}n}} \bibnamefont{and}
  \bibinfo{author}{\bibfnamefont{S.}~\bibnamefont{Durukano\v{g}lu}},
     \textit{\bibinfo{journal}{Phys. Rev. B}} \textbf{\bibinfo{volume}{83}}, \bibinfo{pages}{113409} {\bibinfo{year}{(2011)}}.

\bibitem[{\citenamefont{Velasco}(2011)\citenamefont{}}]{theor8}
  \bibinfo{author}{\bibfnamefont{D.}~\bibnamefont{Mart\'{i}nez-Guti\'{e}rrez}} \bibnamefont{and}
  \bibinfo{author}{\bibfnamefont{V.R.}~\bibnamefont{Velasco}},
     \textit{\bibinfo{journal}{Surf. Sci.}} \textbf{\bibinfo{volume}{605}}, \bibinfo{pages}{24} {\bibinfo{year}{(2011)}}.

\bibitem[{\citenamefont{Mizuno}(2014)\citenamefont{}}]{theor10}
  \bibinfo{author}{\bibfnamefont{S.}~\bibnamefont{Mizuno}},
     \textit{\bibinfo{journal}{Jpn. J. Appl. Phys.}} \textbf{\bibinfo{volume}{53}}, \bibinfo{pages}{07KB02} {\bibinfo{year}{(2014)}Bessas}.

\bibitem[{\citenamefont{Saviot}(2018)\citenamefont{}}]{theor11}
  \bibinfo{author}{\bibfnamefont{L.}~\bibnamefont{Saviot}},
     \textit{\bibinfo{journal}{Phys. Rev. B}} \textbf{\bibinfo{volume}{97}}, \bibinfo{pages}{155420} {\bibinfo{year}{(2018)}}.

\bibitem[{\citenamefont{Zou}(2001)\citenamefont{}}]{therm0}
  \bibinfo{author}{\bibfnamefont{J.}~\bibnamefont{Zou}} \bibnamefont{and}
  \bibinfo{author}{\bibfnamefont{A.}~\bibnamefont{Balandin}}, 
     \textit{\bibinfo{journal}{J. Appl. Phys.}} \textbf{\bibinfo{volume}{89}}, \bibinfo{pages}{2932} {\bibinfo{year}{(2001)}}.

\bibitem[{\citenamefont{Glavin}(2001)\citenamefont{}}]{therm1}
  \bibinfo{author}{\bibfnamefont{B.A.}~\bibnamefont{Glavin}},
     \textit{\bibinfo{journal}{Phys. Rev. Lett.}} \textbf{\bibinfo{volume}{86}}, \bibinfo{pages}{4318} {\bibinfo{year}{(2001)}}.

\bibitem[{\citenamefont{L\"{u}}(2006)\citenamefont{}}]{therm3}
  \bibinfo{author}{\bibfnamefont{X.}~\bibnamefont{L\"{u}}} \bibnamefont{and}
  \bibinfo{author}{\bibfnamefont{J.H.}~\bibnamefont{Chu}}, 
     \textit{\bibinfo{journal}{J. Appl. Phys.}}, \textbf{\bibinfo{volume}{100}}, \bibinfo{pages}{014305} {\bibinfo{year}{(2006)}}.

\bibitem[{\citenamefont{Qiu}(2011)\citenamefont{}}]{therm4}
  \bibinfo{author}{\bibfnamefont{B.}~\bibnamefont{Qiu}},
  \bibinfo{author}{\bibfnamefont{L.}~\bibnamefont{Sun}}, \bibnamefont{and}
  \bibinfo{author}{\bibfnamefont{X.}~\bibnamefont{Ruan}}, 
     \textit{\bibinfo{journal}{Phys. Rev. B}} \textbf{\bibinfo{volume}{83}}, \bibinfo{pages}{035312} {\bibinfo{year}{(2011)}}.

\bibitem[{\citenamefont{Karamitaheri}(2014)\citenamefont{}}]{therm7}
  \bibinfo{author}{\bibfnamefont{H.}~\bibnamefont{Karamitaheri}},
  \bibinfo{author}{\bibfnamefont{N.}~\bibnamefont{Neophytou}}, \bibnamefont{and}
  \bibinfo{author}{\bibfnamefont{H.}~\bibnamefont{Kosina}},
     \textit{\bibinfo{journal}{J. Appl. Phys.}} \textbf{\bibinfo{volume}{115}}, \bibinfo{pages}{024302} {\bibinfo{year}{(2014)}}.

\bibitem[{\citenamefont{Zhou}(2017)\citenamefont{}}]{therm8}
  \bibinfo{author}{\bibfnamefont{Y.}~\bibnamefont{Zhou}},
  \bibinfo{author}{\bibfnamefont{X.}~\bibnamefont{Zhang}}, \bibnamefont{and}
  \bibinfo{author}{\bibfnamefont{M.}~\bibnamefont{Hu}},
     \textit{\bibinfo{journal}{Nano Lett.}} \textbf{\bibinfo{volume}{17}}, \bibinfo{pages}{1269} {\bibinfo{year}{(2017)}}.

\bibitem[{\citenamefont{Rashid}(2018)\citenamefont{}}]{therm10}
  \bibinfo{author}{\bibfnamefont{Z.}~\bibnamefont{Rashid}},
  \bibinfo{author}{\bibfnamefont{L.}~\bibnamefont{Zhu}}, \bibnamefont{and}
  \bibinfo{author}{\bibfnamefont{Wu}~\bibnamefont{Li}},
     \textit{\bibinfo{journal}{Phys. Rev. B}} \textbf{\bibinfo{volume}{97}}, \bibinfo{pages}{075441} {\bibinfo{year}{(2018)}}.

\bibitem[{\citenamefont{Qiu}(2011)\citenamefont{}}]{therm12}
  \bibinfo{author}{\bibfnamefont{B.}~\bibnamefont{Qiu}} \bibnamefont{and}
  \bibinfo{author}{\bibfnamefont{X.}~\bibnamefont{Ruan}}, 
     \textit{\bibinfo{journal}{Phys. Rev. B}} \textbf{\bibinfo{volume}{83}}, \bibinfo{pages}{035312} {\bibinfo{year}{(2011)}}.

\bibitem[{\citenamefont{Nishiguchi}(2002)\citenamefont{}}]{eph1}
  \bibinfo{author}{\bibfnamefont{N.}~\bibnamefont{Nishiguchi}},
     \textit{\bibinfo{journal}{Physica E}} \textbf{\bibinfo{volume}{13}}, \bibinfo{pages}{1} {\bibinfo{year}{(2002)}}.

\bibitem[{\citenamefont{Uno}(2011)\citenamefont{}}]{eph5}
  \bibinfo{author}{\bibfnamefont{S.}~\bibnamefont{Uno}},
  \bibinfo{author}{\bibfnamefont{J.}~\bibnamefont{Hattori}},
  \bibinfo{author}{\bibfnamefont{K.}~\bibnamefont{Nakazato}}, \bibnamefont{and}
  \bibinfo{author}{\bibfnamefont{M.}~\bibnamefont{Nobuya}},
     \textit{\bibinfo{journal}{J. Comput. Electron.}} \textbf{\bibinfo{volume}{10}}, \bibinfo{pages}{104} {\bibinfo{year}{(2011)}}.

\bibitem[{\citenamefont{Yamada}(2012)\citenamefont{}}]{eph6}
  \bibinfo{author}{\bibfnamefont{Y.}~\bibnamefont{Yamada}},
  \bibinfo{author}{\bibfnamefont{H.}~\bibnamefont{Tsuchiya}}, \bibnamefont{and}
  \bibinfo{author}{\bibfnamefont{M.}~\bibnamefont{Ogawa}},
     \textit{\bibinfo{journal}{J. Appl. Phys.}} \textbf{\bibinfo{volume}{111}}, \bibinfo{pages}{063720} {\bibinfo{year}{(2012)}}.

\bibitem[{\citenamefont{Tienda-Luna}(2013)\citenamefont{}}]{eph7}
  \bibinfo{author}{\bibfnamefont{I.M.}~\bibnamefont{Tienda-Luna}},
  \bibinfo{author}{\bibfnamefont{F.G.}~\bibnamefont{Ruiz}},
  \bibinfo{author}{\bibfnamefont{A.}~\bibnamefont{Godoy}},
  \bibinfo{author}{\bibfnamefont{L.}~\bibnamefont{Donetti}},
  \bibinfo{author}{\bibfnamefont{C.}~\bibnamefont{Mart\'{i}nez-Blanque}}, \bibnamefont{and}
  \bibinfo{author}{\bibfnamefont{F.}~\bibnamefont{G\'{a}miz}},
     \textit{\bibinfo{journal}{Appl. Phys. Lett.}} \textbf{\bibinfo{volume}{103}}, \bibinfo{pages}{163107} {\bibinfo{year}{(2013)}}.

\bibitem[{\citenamefont{Malhotra}(2019)\citenamefont{}}]{maldovan3}
  \bibinfo{author}{\bibfnamefont{A.}~\bibnamefont{Malhotra}} \bibnamefont{and}
  \bibinfo{author}{\bibfnamefont{M.}~\bibnamefont{Maldovan}},
     \textit{\bibinfo{journal}{Nanotechnology}} \textbf{\bibinfo{volume}{30}}, \bibinfo{pages}{372002} {\bibinfo{year}{(2019)}}.

\bibitem[{\citenamefont{Li et~al.}(1999)\citenamefont{}}]{zhang}
\bibinfo{author}{\bibfnamefont{B.}~\bibnamefont{Li}},
\bibinfo{author}{\bibfnamefont{D.}~\bibnamefont{Yu}}, \bibnamefont{and}
  \bibinfo{author}{\bibfnamefont{S.-L.}~\bibnamefont{Zhang}},
     \textit{\bibinfo{journal}{Phys. Rev. B}} \textbf{\bibinfo{volume}{59}}, \bibinfo{pages}{1645} {\bibinfo{year}{(1999)}}.

\bibitem[{\citenamefont{Wang et~al.}(2000)\citenamefont{}}]{wang}
\bibinfo{author}{\bibfnamefont{R.P.}~\bibnamefont{Wang}},
\bibinfo{author}{\bibfnamefont{G.W.}~\bibnamefont{Zhou}},
\bibinfo{author}{\bibfnamefont{Y.L.}~\bibnamefont{Liu}},
\bibinfo{author}{\bibfnamefont{S.H.}~\bibnamefont{Pan}},
\bibinfo{author}{\bibfnamefont{H.Z.}~\bibnamefont{Zhang}},
\bibinfo{author}{\bibfnamefont{D.P.}~\bibnamefont{Yu}}, \bibnamefont{and}
  \bibinfo{author}{\bibfnamefont{Z.}~\bibnamefont{Zhang}},
     \textit{\bibinfo{journal}{Phys. Rev. B}} \textbf{\bibinfo{volume}{61}}, \bibinfo{pages}{16827} {\bibinfo{year}{(2000)}}.

\bibitem[{\citenamefont{Piscanec et~al.}(2003)\citenamefont{}}]{piscanec}
\bibinfo{author}{\bibfnamefont{S.}~\bibnamefont{Piscanec}},
\bibinfo{author}{\bibfnamefont{M.}~\bibnamefont{Cantoro}},
\bibinfo{author}{\bibfnamefont{A.C.}~\bibnamefont{Ferrari}},
\bibinfo{author}{\bibfnamefont{J.A.}~\bibnamefont{Zapien}},
\bibinfo{author}{\bibfnamefont{Y.}~\bibnamefont{Lifshitz}},
\bibinfo{author}{\bibfnamefont{S.T.}~\bibnamefont{Lee}},
\bibinfo{author}{\bibfnamefont{S.}~\bibnamefont{Hofmann}}, \bibnamefont{and}
  \bibinfo{author}{\bibfnamefont{J.}~\bibnamefont{Robertson}},
     \textit{\bibinfo{journal}{Phys. Rev. B}} \textbf{\bibinfo{volume}{68}}, \bibinfo{pages}{241312(R)} {\bibinfo{year}{(2003)}}.

\bibitem[{\citenamefont{Adu et~al.}(2005)\citenamefont{}}]{adu}
  \bibinfo{author}{\bibfnamefont{K.W.}~\bibnamefont{Adu}},
  \bibinfo{author}{\bibfnamefont{H.R.}~\bibnamefont{Gutierrez}},
  \bibinfo{author}{\bibfnamefont{U.J.}~\bibnamefont{Kim}},
\bibinfo{author}{\bibfnamefont{G.U.}~\bibnamefont{Sumanasekera}}, \bibnamefont{and}
  \bibinfo{author}{\bibfnamefont{P.C.}~\bibnamefont{Eklund}},
     \textit{\bibinfo{journal}{Nano Lett.}} \textbf{\bibinfo{volume}{5}}, \bibinfo{pages}{409} {\bibinfo{year}{(2005)}}.

\bibitem[{\citenamefont{Patsha et~al.}(2018)\citenamefont{}}]{dhara}
\bibinfo{author}{\bibfnamefont{A.}~\bibnamefont{Patsha}} \bibnamefont{and}
  \bibinfo{author}{\bibfnamefont{S.}~\bibnamefont{Dhara}},
     \textit{\bibinfo{journal}{Nano Lett.}} \textbf{\bibinfo{volume}{18}}, \bibinfo{pages}{7181} {\bibinfo{year}{(2018)}}.

\bibitem[{\citenamefont{Luca et~al.}(2019)\citenamefont{}}]{luca}
  \bibinfo{author}{\bibfnamefont{M.}~\bibnamefont{De Luca}},
  \bibinfo{author}{\bibfnamefont{C.}~\bibnamefont{Fasolato}},
  \bibinfo{author}{\bibfnamefont{M.A.}~\bibnamefont{Verheijen}},
  \bibinfo{author}{\bibfnamefont{Y.}~\bibnamefont{Ren}},
  \bibinfo{author}{\bibfnamefont{M.Y.}~\bibnamefont{Swinkels}},
  \bibinfo{author}{\bibfnamefont{S.}~\bibnamefont{K\"{o}lling}},
  \bibinfo{author}{\bibfnamefont{E.P.A.M.}~\bibnamefont{Bakkers}},
  \bibinfo{author}{\bibfnamefont{R.}~\bibnamefont{Rurali}},
\bibinfo{author}{\bibfnamefont{X.}~\bibnamefont{Cartoixa}}, \bibnamefont{and}
  \bibinfo{author}{\bibfnamefont{I.}~\bibnamefont{Zardo}},
     \textit{\bibinfo{journal}{Nano Lett.}} \textbf{\bibinfo{volume}{19}}, \bibinfo{pages}{4702} {\bibinfo{year}{(2019)}}.

\bibitem[{\citenamefont{Mante et~al.}(2018)\citenamefont{}}]{perrin}
  \bibinfo{author}{\bibfnamefont{P.-A.}~\bibnamefont{Mante}},
\bibinfo{author}{\bibfnamefont{L.}~\bibnamefont{Belliard}}, \bibnamefont{and}
  \bibinfo{author}{\bibfnamefont{B.}~\bibnamefont{Perrin}},
     \textit{\bibinfo{journal}{Nanophotonics}} \textbf{\bibinfo{volume}{7}}, \bibinfo{pages}{1759} {\bibinfo{year}{(2018)}}.

\bibitem[{\citenamefont{Mariager et~al.}(2010)\citenamefont{}}]{mariager}
  \bibinfo{author}{\bibfnamefont{A.O.}~\bibnamefont{Mariager}},
  \bibinfo{author}{\bibfnamefont{D.}~\bibnamefont{Khakhulin}},
  \bibinfo{author}{\bibfnamefont{H.T.}~\bibnamefont{Lemke}},
  \bibinfo{author}{\bibfnamefont{K.S.}~\bibnamefont{Kjaer}},
  \bibinfo{author}{\bibfnamefont{L.}~\bibnamefont{Guerin}},
  \bibinfo{author}{\bibfnamefont{L.}~\bibnamefont{Nuccio}},
   \bibinfo{author}{\bibfnamefont{C.B.}~\bibnamefont{S{\o}rensen}},
  \bibinfo{author}{\bibfnamefont{M.N.}~\bibnamefont{Nielsen}}, \bibnamefont{and}
\bibinfo{author}{\bibfnamefont{R.}~\bibnamefont{Feidenhans'l}},
     \textit{\bibinfo{journal}{Nano Lett.}} \textbf{\bibinfo{volume}{10}}, \bibinfo{pages}{2461} {\bibinfo{year}{(2010)}}.

\bibitem[{\citenamefont{Kargar et~al.}(2016)\citenamefont{}}]{balandin3}
\bibinfo{author}{\bibfnamefont{F.}~\bibnamefont{Kargar}},
\bibinfo{author}{\bibfnamefont{B.}~\bibnamefont{Debnath}},
\bibinfo{author}{\bibfnamefont{J.-P.}~\bibnamefont{Kakko}},
\bibinfo{author}{\bibfnamefont{A.}~\bibnamefont{S{\"a}yn{\"a}tjoki}},
\bibinfo{author}{\bibfnamefont{H.}~\bibnamefont{Lipsanen}},
\bibinfo{author}{\bibfnamefont{D.L.}~\bibnamefont{Nika}},
\bibinfo{author}{\bibfnamefont{R.K.}~\bibnamefont{Lake}}, \bibnamefont{and}
  \bibinfo{author}{\bibfnamefont{A.A.}~\bibnamefont{Balandin}},
     \textit{\bibinfo{journal}{Nature Com.}} \textbf{\bibinfo{volume}{7}}, \bibinfo{pages}{13400} {\bibinfo{year}{(2016)}}.

\bibitem[{\citenamefont{Bessas et~al.}(2013)\citenamefont{}}]{bessas}
\bibinfo{author}{\bibfnamefont{D.}~\bibnamefont{Bessas}},
\bibinfo{author}{\bibfnamefont{W.}~\bibnamefont{T{\"o}llner}},
\bibinfo{author}{\bibfnamefont{Z.}~\bibnamefont{Aabdin}},
\bibinfo{author}{\bibfnamefont{N.}~\bibnamefont{Peranio}},
\bibinfo{author}{\bibfnamefont{I.}~\bibnamefont{Sergueev}},
\bibinfo{author}{\bibfnamefont{H.-C.}~\bibnamefont{Wille}},
\bibinfo{author}{\bibfnamefont{O.}~\bibnamefont{Eibl}},
  \bibinfo{author}{\bibfnamefont{K.}~\bibnamefont{Nielsch}}, \bibnamefont{and}
  \bibinfo{author}{\bibfnamefont{R.P.}~\bibnamefont{Hermann}},
     \textit{\bibinfo{journal}{Nanoscale}} \textbf{\bibinfo{volume}{5}}, \bibinfo{pages}{10629} {\bibinfo{year}{(2013)}}.

\bibitem[{\citenamefont{Temst{\"o}gl et~al.}(2019)\citenamefont{}}]{temst}
\bibinfo{author}{\bibfnamefont{D.P.}~\bibnamefont{Lozano}},
\bibinfo{author}{\bibfnamefont{S.}~\bibnamefont{Couet}},
\bibinfo{author}{\bibfnamefont{C.}~\bibnamefont{Petermann}},
\bibinfo{author}{\bibfnamefont{G.}~\bibnamefont{Hamoir}},
\bibinfo{author}{\bibfnamefont{J.K.}~\bibnamefont{Jochum}},
\bibinfo{author}{\bibfnamefont{T.}~\bibnamefont{Picot}},
\bibinfo{author}{\bibfnamefont{E.}~\bibnamefont{Menendez}},
\bibinfo{author}{\bibfnamefont{K.}~\bibnamefont{Houben}},
\bibinfo{author}{\bibfnamefont{V.}~\bibnamefont{Joly}},
\bibinfo{author}{\bibfnamefont{V.A.}~\bibnamefont{Antohe}},
\bibinfo{author}{\bibfnamefont{M.Y.}~\bibnamefont{Hu}},
\bibinfo{author}{\bibfnamefont{B.M.}~\bibnamefont{Leu}},
\bibinfo{author}{\bibfnamefont{A.}~\bibnamefont{Alatas}},
\bibinfo{author}{\bibfnamefont{A.H.}~\bibnamefont{Said}},
\bibinfo{author}{\bibfnamefont{S.}~\bibnamefont{Roelants}},
\bibinfo{author}{\bibfnamefont{B.}~\bibnamefont{Partoens}},
\bibinfo{author}{\bibfnamefont{M.V.}~\bibnamefont{Milosevic}},
\bibinfo{author}{\bibfnamefont{F.M.}~\bibnamefont{Peeters}},
  \bibinfo{author}{\bibfnamefont{L.}~\bibnamefont{Piraux}}, 
  \bibinfo{author}{\bibfnamefont{J.}~\bibnamefont{Van de Vondel}},
  \bibinfo{author}{\bibfnamefont{A.}~\bibnamefont{Vantomme}},
  \bibinfo{author}{\bibfnamefont{K.}~\bibnamefont{Temst}}, \bibnamefont{and}
  \bibinfo{author}{\bibfnamefont{M.J.}~\bibnamefont{Van Bael}},
     \textit{\bibinfo{journal}{Phys. Rev. B}} \textbf{\bibinfo{volume}{99}}, \bibinfo{pages}{064512} {\bibinfo{year}{(2019)}}.

\bibitem[{\citenamefont{Krause et~al.}(2012)\citenamefont{}}]{sputter_krause}
   \bibinfo{author}{\bibfnamefont{B.}~\bibnamefont{Krause}},
   \bibinfo{author}{\bibfnamefont{S.}~\bibnamefont{Darma}},    
   \bibinfo{author}{\bibfnamefont{M.}~\bibnamefont{Kaufholz}},
   \bibinfo{author}{\bibfnamefont{H.-H.}~\bibnamefont{Gr\"afe}},    
   \bibinfo{author}{\bibfnamefont{S.}~\bibnamefont{Ulrich}},    
  \bibinfo{author}{\bibfnamefont{M.}~\bibnamefont{Mantilla}},     
   \bibinfo{author}{\bibfnamefont{R.}~\bibnamefont{Weigel}},     
  \bibinfo{author}{\bibfnamefont{S.}~\bibnamefont{Rembold}}, \bibnamefont{and}        
   \bibinfo{author}{\bibfnamefont{T.}~\bibnamefont{Baumbach}},    
  \textit{\bibinfo{journal}{J. Synchrotron Radiat.}} \textbf{\bibinfo{volume}{19}}, \bibinfo{pages}{216} \bibinfo{year}{(2012)}.
    
\bibitem[{\citenamefont{Ravel et~al.}(2005)}]{ravel_athena}
\bibinfo{author}{\bibfnamefont{B.} \bibnamefont{Ravel}} \bibnamefont{and} 
\bibinfo{author}{\bibfnamefont{M.} \bibnamefont{Newville}},
 \textit{ \bibinfo{journal}{J. Synchrotron Radiat.}},
 \textbf{\bibinfo{volume}{12}} \bibinfo{pages}{537} \bibinfo{year}{(2005}.

\bibitem[{\citenamefont{Kohn and Chumakov}(2000)}]{evaluation_pdos}
\bibinfo{author}{\bibfnamefont{V.G.} \bibnamefont{Kohn}} \bibnamefont{and}
  \bibinfo{author}{\bibfnamefont{A.I.} \bibnamefont{Chumakov}},
  \textit{\bibinfo{journal}{Hyperfine Interact.}} \textbf{\bibinfo{volume}{125}},   \bibinfo{pages}{205} \bibinfo{year}{(2000)}.
  
  
\bibitem{Seto_PRL_NIS} M. Seto, Y. Yoda, S. Kikuta, X.W. Zhang, and M. Ando, \textit{Phys. Rev. Lett.} \textbf{74}, 3828 (1995).

\bibitem{Sturhahn_PRL_NIS} W. Sturhahn, T.S. Toellner, E.E. Alp, X. Zhang, M. Ando, Y. Yoda and S. Kikuta, M. Seto, C.W. Kimball, and B. Dabrowski , \textit{Phys. Rev. Lett.} \textbf{74}, 3832 (1995).

    
\bibitem[{\citenamefont{Wille et~al.}(2010)\citenamefont{Wille, Franz,
  R{\"o}hlsberger, Caliebe, and Dill}}]{p01}
\bibinfo{author}{\bibfnamefont{H.-C.} \bibnamefont{Wille}},
  \bibinfo{author}{\bibfnamefont{H.}~\bibnamefont{Franz}},
  \bibinfo{author}{\bibfnamefont{R.}~\bibnamefont{R{\"o}hlsberger}},
  \bibinfo{author}{\bibfnamefont{W.A.} \bibnamefont{Caliebe}},
  \bibnamefont{and} \bibinfo{author}{\bibfnamefont{F.U.} \bibnamefont{Dill}},
  \textit{\bibinfo{journal}{J. Phys.: Conf. Series}}
 \textbf{\bibinfo{volume}{217}},    \bibinfo{pages}{0120081}  \bibinfo{year}{(2010)}.
  
\bibitem[{\citenamefont{R{\"u}ffer and Chumakov}(1996)\citenamefont{R{\"u}ffer, Chumakov}}]{id18}
\bibinfo{author}{\bibfnamefont{R.} \bibnamefont{R{\"u}ffer}}
  \bibnamefont{and} \bibinfo{author}{\bibfnamefont{A.I.} \bibnamefont{Chumakov}},
  \textit{\bibinfo{journal}{Hyperfine Interact.}}
 \textbf{\bibinfo{volume}{97}}, \bibinfo{pages}{589}  \bibinfo{year}{(1996)}.
  
\bibitem[{\citenamefont{Ibrahimkutty et~al.}(2015)}]{ibrahimkutty_chamber}
\bibinfo{author}{\bibfnamefont{S.} \bibnamefont{Ibrahimkutty}},
  \bibinfo{author}{\bibfnamefont{A.}~\bibnamefont{Seiler}},
  \bibinfo{author}{\bibfnamefont{T.}~\bibnamefont{Pr{\"u}{\ss}mann}},
   \bibinfo{author}{\bibfnamefont{T.}~\bibnamefont{Vitova}},
   \bibinfo{author}{\bibfnamefont{R.}~\bibnamefont{Pradip}},
   \bibinfo{author}{\bibfnamefont{O.}~\bibnamefont{Bauder}},   
   \bibinfo{author}{\bibfnamefont{P.}~\bibnamefont{Wochner}},      
   \bibinfo{author}{\bibfnamefont{A.}~\bibnamefont{Plech}},      
   \bibinfo{author}{\bibfnamefont{T.}~\bibnamefont{Baumbach}},   
  \bibnamefont{and} \bibinfo{author}{\bibfnamefont{S.} \bibnamefont{Stankov}},
  \textit{\bibinfo{journal}{J. Synchrotron Radiat.}}
 \textbf{\bibinfo{volume}{22}},  \bibinfo{pages}{91}  \bibinfo{year}{(2015)}.
  
\bibitem{vasp1} G. Kresse and J. Furthm\"{u}ller, \textit{Phys. Rev. B}  \textbf{54}, 11169 (1996).

\bibitem{vasp2} G. Kresse and J. Furthm\"{u}ller, \textit{Comput. Mater. Sci.}  \textbf{6}, 15 (1996).

\bibitem{PBE1} J.P. Perdew, K. Burke, and M. Ernzerhof, \textit{Phys. Rev. Lett.}  \textbf{77}, 3865 (1996).

\bibitem{PBE2} J.P. Perdew, K. Burke, and M. Ernzerhof, \textit{Phys. Rev. Lett.} \textbf{78}, 1396 (1997).

\bibitem{phonon1} K. Parlinski, Z.Q. Li, and Y. Kawazoe, \textit{Phys. Rev. Lett.} \textbf{78}, 4063 (1997).

\bibitem{phonon2} K. Parlinski, Software PHONON ver. 6.15, Cracow, Poland, 2015.

\bibitem[{\citenamefont{Kalt et~al.}(2020)\citenamefont{}}]{kalt_alpha_islands}
  \bibinfo{author}{\bibfnamefont{J.}~\bibnamefont{Kalt}},
   \bibinfo{author}{\bibfnamefont{M.}~\bibnamefont{Sternik}}, 
   \bibinfo{author}{\bibfnamefont{B.}~\bibnamefont{Krause}}, 
   \bibinfo{author}{\bibfnamefont{I.}~\bibnamefont{Sergueev}},   
   \bibinfo{author}{\bibfnamefont{M.}~\bibnamefont{Mikolasek}},   
   \bibinfo{author}{\bibfnamefont{D.}~\bibnamefont{Bessas}},   
   \bibinfo{author}{\bibfnamefont{O.}~\bibnamefont{Sikora}},   
   \bibinfo{author}{\bibfnamefont{T.}~\bibnamefont{Vitova}},   
   \bibinfo{author}{\bibfnamefont{J.}~\bibnamefont{G\"ottlicher}},   
   \bibinfo{author}{\bibfnamefont{R.}~\bibnamefont{Steininger}},   
   \bibinfo{author}{\bibfnamefont{P.T.}~\bibnamefont{Jochym}},   
   \bibinfo{author}{\bibfnamefont{A.}~\bibnamefont{Ptok}},   
   \bibinfo{author}{\bibfnamefont{O.}~\bibnamefont{Leupold}},   
   \bibinfo{author}{\bibfnamefont{H.-C.}~\bibnamefont{Wille}},   
   \bibinfo{author}{\bibfnamefont{A.I.}~\bibnamefont{Chumakov}},   
  \bibinfo{author}{\bibfnamefont{P.}~\bibnamefont{Piekarz}},
  \bibinfo{author}{\bibfnamefont{K.}~\bibnamefont{Parlinski}},    
  \bibinfo{author}{\bibfnamefont{T.}~\bibnamefont{Baumbach}}, \bibnamefont{and}            
   \bibinfo{author}{\bibfnamefont{S.}~\bibnamefont{Stankov}},    
  \textit{\bibinfo{journal}{Phys. Rev. B}} \textbf{\bibinfo{volume}{101}}, \bibinfo{pages}{165406} \bibinfo{year}{(2020)}.

\bibitem{1d_RHEED_theo} P. Delescluse and A. Masson, \textit{Surf. Sci.} \textbf{100}, 423 (1980).

\bibitem{dobson1982_GaAs_domain} P. J. Dobson,  J.H. Neave and B.A. Joyce, \textit{Surf. Sci.} \textbf{119}, L339 (1982).

\bibitem{1d_RHEED_exp} G. Wang, S.K. Lok, S.K. Chan, C. Wang, G.K.L. Wong and I.K. Sou, \textit{Nanotechnology}  \textbf{20}, 215607 (2009).

\bibitem{sm} See Supplemental Material at http for RHEED, EXAFS, AFM data evaluation, calculation of the direction-projected phonon density of states, and comparison of \textit{ex situ} and \textit{in situ} PDOS.

\bibitem[{\citenamefont{Berbezier et~al.}(1994)\citenamefont{}}]{berbezier_TEM}
\bibinfo{author}{\bibfnamefont{I.}~\bibnamefont{Berbezier}},
  \bibinfo{author}{\bibfnamefont{J.}~\bibnamefont{Chevrier}}, \bibnamefont{and}
  \bibinfo{author}{\bibfnamefont{J.}~\bibnamefont{Derrien}},
  \textit{\bibinfo{journal}{Surf. Sci.}} \textbf{\bibinfo{volume}{315}}, \bibinfo{pages}{27} \bibinfo{year}{(1994)}.
  
\bibitem[{\citenamefont{Kataoka et~al.}(2006)\citenamefont{}}]{kataoka_phase}
  \bibinfo{author}{\bibfnamefont{K.}~\bibnamefont{Kataoka}},
  \bibinfo{author}{\bibfnamefont{K.}~\bibnamefont{Hattori}},
  \bibinfo{author}{\bibfnamefont{Y.}~\bibnamefont{Miyatake}}, \bibnamefont{and}  
  \bibinfo{author}{\bibfnamefont{H.}~\bibnamefont{Daimon}},
  \textit{\bibinfo{journal}{Phys. Rev. B}} \textbf{\bibinfo{volume}{74}}, \bibinfo{pages}{155406} \bibinfo{year}{(2006)}.


  \bibitem[{\citenamefont{Chumakov and Sturhahn}(1999)}]{NIS_selctivity}
\bibinfo{author}{\bibfnamefont{A.I.} \bibnamefont{Chumakov}} \bibnamefont{and}
  \bibinfo{author}{\bibfnamefont{W.} \bibnamefont{Sturhahn}},
  \textit{\bibinfo{journal}{Hyperfine Interact.}} \textbf{\bibinfo{volume}{123/124}},  \bibinfo{pages}{718} \bibinfo{year}{(1999)}.
  
    
\bibitem{chen_alpha_wires} S.Y. Chen, H.C. Chen, and L.J. Chen, \textit{Appl. Phys. Lett.} \textbf{88} 193114 (2006).

\bibitem{won_alpha_2006}  J.H. Won, K. Sato, M. Ishimaru, and Y. Hirotsu, \textit{J. Appl. Phys.} \textbf{100}, 014307 (2006).
  
  
\bibitem[{\citenamefont{Chumakov et al.}(1997)}]{chumakov_anisotropic_nis_FeBO3}
\bibinfo{author}{\bibfnamefont{A.I.} \bibnamefont{Chumakov}},
  \bibinfo{author}{\bibfnamefont{R.} \bibnamefont{R{\"u}ffer}},
  \bibinfo{author}{\bibfnamefont{A.Q.R.} \bibnamefont{Baron}},
  \bibinfo{author}{\bibfnamefont{H.} \bibnamefont{Gr{\"u}nsteudel}},
  \bibinfo{author}{\bibfnamefont{H.F.} \bibnamefont{Gr{\"u}nsteudel}}, \bibnamefont{and}
  \bibinfo{author}{\bibfnamefont{V.G.} \bibnamefont{Kohn}},
 \textit{\bibinfo{journal}{Phys. Rev. B}} \textbf{\bibinfo{volume}{56}},
  \bibinfo{pages}{10758} (\bibinfo{year}{1997)}.

\bibitem[{\citenamefont{Kohn et al.}(1998)}]{kohn_anisotropic_nis_theo}
\bibinfo{author}{\bibfnamefont{V.G.} \bibnamefont{Kohn}},
	\bibinfo{author}{\bibfnamefont{A.I.} \bibnamefont{Chumakov}}, \bibnamefont{and}
	\bibinfo{author}{\bibfnamefont{R.} \bibnamefont{R{\"u}ffer}},
    \textit{\bibinfo{journal}{Phys. Rev. B}} \textbf{\bibinfo{volume}{58}},
    \bibinfo{pages}{8437} (\bibinfo{year}{1998}).

\bibitem[{\citenamefont{Fultz}(2010)}]{fultz_dho}
\bibinfo{author}{\bibfnamefont{B.}~\bibnamefont{Fultz}},
  \textit{\bibinfo{journal}{Prog. Mater. Sci.}}  \textbf{\bibinfo{volume}{55}}, \bibinfo{pages}{247} \bibinfo{year}{(2010)}.
  
\bibitem{Faak1}  B. F\aa k and B. Dorner, Institute Laue  Langevin Technical Report No. {92FA008S}, 1992, (unpublished); B. F\aa k and B. Dorner, \textit{Physica B} \textbf{234}, 1107 (1997).

\bibitem{note_Axy_Az} The experimental data was additionally modeled with \textit{A$_{xy}$} and \textit{A$_z$} being free parameters in the least squares optimization.
The quality factors obtained with this approach coincide within the uncertainty with the values given in Fig. \ref{fig:PDOS}, \textit{A$_{xy}$} and \textit{A$_z$} deviate at most by 2\,\% from the calculated values of \textit{A$_{xy}$}=0.9 and \textit{A$_z$}=0.1. 

\bibitem[{\citenamefont{Pradip et~al.}(2019)\citenamefont{}}]{pradip_EuO_nscale}
  \bibinfo{author}{\bibfnamefont{R.}~\bibnamefont{Pradip}}, 
  \bibinfo{author}{\bibfnamefont{P.}~\bibnamefont{Piekarz}},
  \bibinfo{author}{\bibfnamefont{D.G.}~\bibnamefont{Merkel}},
  \bibinfo{author}{\bibfnamefont{J.}~\bibnamefont{Kalt}},
  \bibinfo{author}{\bibfnamefont{O.}~\bibnamefont{Waller}},
  \bibinfo{author}{\bibfnamefont{A.I.}~\bibnamefont{Chumakov}},    
  \bibinfo{author}{\bibfnamefont{R.}~\bibnamefont{R{\"u}ffer}},    
  \bibinfo{author}{\bibfnamefont{A.M.}~\bibnamefont{Ole{\'s}}},
  \bibinfo{author}{\bibfnamefont{K.}~\bibnamefont{Parlinski}},       
  \bibinfo{author}{\bibfnamefont{T.}~\bibnamefont{Baumbach}}, \bibnamefont{and}      
   \bibinfo{author}{\bibfnamefont{S.}~\bibnamefont{Stankov}},  
  \textit{\bibinfo{journal}{Nanoscale}} \textbf{\bibinfo{volume}{116}},
  \bibinfo{pages}{10968} (\bibinfo{year}{2019}). 

\bibitem{Fe3Si_interface} J. Kalt, M. Sternik, I. Sergueev, J. Herfort, B. Jenichen, H.-C. Wille, O. Sikora, P. Piekarz, K. Parlinski, T. Baumbach, and S. Stankov, \textit{Phys. Rev. B} \textbf{98}, 121409(R) (2018).


\bibitem[{\citenamefont{{\'S}lezak et~al.}(2007)\citenamefont{{\'S}lezak,
  {\L}a{\.z}ewski, Stankov, Parlinski, Reitinger, Rennhofer, R{\"u}ffer,
  Sepiol, {\'S}lezak, and Spiridis}}]{slezak1}
\bibinfo{author}{\bibfnamefont{T.}~\bibnamefont{{\'S}lezak}},
  \bibinfo{author}{\bibfnamefont{J.}~\bibnamefont{{\L}a{\.z}ewski}},
  \bibinfo{author}{\bibfnamefont{S.}~\bibnamefont{Stankov}},
  \bibinfo{author}{\bibfnamefont{K.}~\bibnamefont{Parlinski}},
  \bibinfo{author}{\bibfnamefont{R.}~\bibnamefont{Reitinger}},
  \bibinfo{author}{\bibfnamefont{M.}~\bibnamefont{Rennhofer}},
  \bibinfo{author}{\bibfnamefont{R.}~\bibnamefont{R{\"u}ffer}},
  \bibinfo{author}{\bibfnamefont{B.}~\bibnamefont{Sepiol}},
  \bibinfo{author}{\bibfnamefont{M.}~\bibnamefont{{\'S}lezak}},
  \bibinfo{author}{\bibfnamefont{N.}~\bibnamefont{Spiridis}},
  \bibinfo{author}{\bibfnamefont{M.}~\bibnamefont{Zajac}},
  \bibinfo{author}{\bibfnamefont{A.I.}~\bibnamefont{Chumakov}}, \bibnamefont{and}
  \bibinfo{author}{\bibfnamefont{J.}~\bibnamefont{Korecki}},
  \textit{\bibinfo{journal}{Phys. Rev. Lett.}}  \textbf{\bibinfo{volume}{99}},  \bibinfo{pages}{0661031} \bibinfo{year}{(2007)}.



\bibitem[{\citenamefont{Stankov et~al.}(2007)\citenamefont{Stankov,
  R{\"o}hlsberger, {\'S}lezak, Sladecek, Sepiol, Vogl, Chumakov, R{\"u}ffer,
  Spiridis, and {\L}a{\.z}ewski}}]{stankov1}
\bibinfo{author}{\bibfnamefont{S.}~\bibnamefont{Stankov}},
  \bibinfo{author}{\bibfnamefont{R.}~\bibnamefont{R{\"o}hlsberger}},
  \bibinfo{author}{\bibfnamefont{T.}~\bibnamefont{{\'S}lezak}},
  \bibinfo{author}{\bibfnamefont{M.}~\bibnamefont{Sladecek}},
  \bibinfo{author}{\bibfnamefont{B.}~\bibnamefont{Sepiol}},
  \bibinfo{author}{\bibfnamefont{G.}~\bibnamefont{Vogl}},
  \bibinfo{author}{\bibfnamefont{A.I.}~\bibnamefont{Chumakov}},
  \bibinfo{author}{\bibfnamefont{R.}~\bibnamefont{R{\"u}ffer}},
  \bibinfo{author}{\bibfnamefont{N.}~\bibnamefont{Spiridis}}, 
  \bibinfo{author}{\bibfnamefont{J.}~\bibnamefont{{\L}a{\.z}ewski}},
  \bibinfo{author}{\bibfnamefont{K.}~\bibnamefont{Parlinski}},\bibnamefont{and}
  \bibinfo{author}{\bibfnamefont{J.}~\bibnamefont{Korecki}},
  \textit{\bibinfo{journal}{Phys. Rev. Lett.}} \textbf{\bibinfo{volume}{99}},  \bibinfo{pages}{1855011} \bibinfo{year}{(2007)}.
  
  
\bibitem[{\citenamefont{Hu et~al.}(2003)\citenamefont{}}]{hu_sos}
   \bibinfo{author}{\bibfnamefont{M.Y.}~\bibnamefont{Hu}},
   \bibinfo{author}{\bibfnamefont{W.}~\bibnamefont{Sturhahn}},   
   \bibinfo{author}{\bibfnamefont{T.S.}~\bibnamefont{Toellner}},     
   \bibinfo{author}{\bibfnamefont{P.D.}~\bibnamefont{Mannheim}},  
   \bibinfo{author}{\bibfnamefont{D.E.}~\bibnamefont{Brown}},       
  \bibinfo{author}{\bibfnamefont{J.}~\bibnamefont{Zhao}},  \bibnamefont{and}        
   \bibinfo{author}{\bibfnamefont{E.E.}~\bibnamefont{Alp}},    
  \textit{\bibinfo{journal}{Phys. Rev. B}} \textbf{\bibinfo{volume}{67}},  \bibinfo{pages}{094304} \bibinfo{year}{(2003)}.
  

\end{thebibliography}
\end{document}